\documentclass[twocolumn,aps,prb,showpacs,floatfix,twoside]{revtex4}

\usepackage{amsmath}
\usepackage{bbm}
\usepackage{graphicx}

\newcommand{\average}[1]{\langle #1\rangle}
\newcommand{\inner}[2]{\langle #1 | #2 \rangle}
\newcommand{\mel}[3]{\langle #1 | #2 | #3\rangle}
\newcommand{\identity}{\mathbbm{1}}

\newcommand{\symbwidth}{0.82ex}
\newcommand{\gy}{}
\newcounter{gx}
\newcommand{\set}[1]{\put(\value{gx},\gy){#1}\addtocounter{gx}{1}}
\newcommand{\nxt}{\renewcommand{\gy}{0}\setcounter{gx}{0}}
\newcommand{\setO}{\set{$\circ$}}
\newcommand{\setX}{\set{$\bullet$}}
\newcommand{\cbar}{\set{\rule{0.13ex}{0ex}\rule[0.5ex]{\symbwidth}{0.2ex}}}
\newcommand{\txt}[1]{\set{\shortstack[l]{\footnotesize $#1$\\\rule{0.13ex}{0ex}\rule[0.5ex]{\symbwidth}{0.2ex}}}}
\newcommand{\ltxt}[1]{\put(\value{gx},\gy){\raisebox{-1ex}{\footnotesize $#1$}}\cbar}
\newcommand{\diagram}[2]{\setlength{\unitlength}{\symbwidth}\begin{picture}(#1,3.7)(0,0.6)\renewcommand{\gy}{1}\setcounter{gx}{0}#2\end{picture}}
\newcommand{\dline}[2]{\diagram{#1}{\renewcommand{\gy}{0.5}#2}}
\newcommand{\textdline}[2]{\setlength{\unitlength}{\symbwidth}\begin{picture}(#1,2.5)(0,.6)\setcounter{gx}{0}\renewcommand{\gy}{0.5}#2\end{picture}}

\begin{document}

\title{Glassy dynamics and domains: exact results for the East model}

\author{Ramses \surname{van Zon}}
\affiliation{The Rockefeller University, 1230 York Avenue, New York,
New York 10021, USA}
\affiliation{Chemical Physics Theory Group, Dept. of Chemistry,
University of Toronto, Ontario, Canada M5S 3H6}

\author{Jeremy Schofield}
\affiliation{Chemical Physics Theory Group, Dept. of Chemistry,
University of Toronto, Ontario, Canada M5S 3H6}

\date{December 8, 2004}

\begin{abstract}
\noindent
A general matrix-based scheme for analyzing the long-time dynamics in
kinetically constrained models such as the East model is presented.
The treatment developed here is motivated by the expectation that
slowly-relaxing spin domains of arbitrary size govern the highly
cooperative events that lead to spin relaxation at long times.  To
account for the role of large spin domains in the dynamics, a complete
basis expressed in terms of domains of all sizes is introduced.  It is
first demonstrated that accounting for single domains of all possible
sizes leads to a simple analytical result for the two-time single-spin
correlation function in the East model that is in excellent
quantitative agreement with simulation data for equilibrium spin up
density values $c \geq 0.6$.  It is then shown that including also two
neighboring domains leads to a closed expression that describes the
slow relaxation of the system down to $c \approx 0.3$.  Ingredients of
generalizing the method to lower values of $c$ are also provided, as
well as to other models.  The main advantage of this approach is that
it gives explicit analytical results and that it requires neither an
arbitrary closure for the memory kernel nor the construction of an
irreducible memory kernel. It also allows one to calculate quantities
that measure heterogeneity in the same framework, as is illustrated on
the neighbor-pair correlation function and the distribution of
relaxation times.
\end{abstract}


\maketitle

\section{\label{introduction}Introduction}

Despite much progress in recent years, many aspects of structural
glasses and undercooled liquids still escape a complete
understanding\cite{Angell95,Ediger96,DebenedettiStillinger01,Ediger00}.
Rather than studying the behavior of molecular glasses, one often
investigates the behavior of simple models in the hope to capture the
basic physics of such systems.  The so-called East model is one of
these simple models\cite{JaeckleEisinger91}.  It is a classic Ising
model with a trivial Hamiltonian in which the stochastic dynamics
governing the change of spin leads to a complicated and
highly-cooperative evolution of the system. In the East model, any
spin has finite probability to flip from up to down or vice-versa only
if the spin to the east of it (i.e., of higher lattice index)
is up. Models of this kind are generally called kinetically
constrained models or facilitating spin
models\cite{FredericksonAndersen84,FredericksonAndersen85,Pittsetal00,PittsAndersen01,RitortSollich03}.

Such models are designed to
mimic\cite{GarrahanChandler02,GarrahanChandler03b,RitortSollich03,SollichEvans03}
the kind of dynamics that take place in glasses\cite{Ediger00}.
Although the East model itself does not have a glass transition at any
finite spin density\cite{JaeckleEisinger91}, the decay of the single
spin time-correlation function at low densities is a stretched
exponential\cite{WuCao04,FredericksonAndersen85}, following a
functional form similar to that of the dynamic structure factor in
glassy systems\cite{Palmeretal84,GoetzeSjoegren88,Phillips96}.  In
fact, the typical spin relaxation time has been shown to behave as
$\log(\tau)\sim \log^{2}(1/c)$, where $c$ is the equilibrium density
of up-spins, heuristically by Sollich and Evans\cite{SollichEvans99}
and rigorously by Aldous and Diaconis\cite{AldousDiaconis02},
indicating an extreme slowing down for small $c$, suggestive of a
transition at $c=0$.  More recently, the East model has been analyzed
to examine the nature of dynamic
heterogeneities\cite{GarrahanChandler02,GarrahanChandler03b} in
frustrated systems.

The typical relaxation times present in systems exhibiting frustration
can be retrieved from time correlation functions. In glasses, mode
coupling theory is one of the predominant descriptions for these
correlation functions.  Several approaches to mode coupling theories
exist. In the context of glasses, that of G\"otze and co-workers has
been widely
used\cite{GotzeLucke75,Bosseetal78a,Bengtzeliusetal84,Leutheusser84,GotzeSjoegren87,GoetzeSjoegren88,Fuchsetal91,GoetzeSjoegren92,Cumminsetal93,Fuchsetal98,Goetze99,GoetzeSperl02}.

The approach of G\"otze and co-workers expresses the time correlation
functions in terms of a memory kernel and then uses a certain ansatz
in which the memory function is written in terms of the correlation
functions themselves, yielding self-consistent equations.  Oppenheim
and co-workers have addressed the formal points and justification of
mode coupling theories in Fourier
space\cite{MachtaOppenheim82,Schofieldetal92,LiuOppenheim97,LiuOppenheim97b}.
Along similar lines, Andersen\cite{Andersenpreprints00} has formulated
a phase-space mode coupling theory for general fluids that leads to
self-consistent equations for time dependent correlation functions.

Mode coupling theory can be applied for both deterministic and
stochastic systems.  For some systems, such as the East model, the
application of the most commonly used mode-coupling ansatz necessary
to close the resulting equation of motion for the spin autocorrelation
function leads to a spurious transition from an ergodic to non-ergodic
phase at finite values of the spin concentration $c$, a result that is
clearly at odds with simulation results.  To analyze the failure of
the closure approximation in mode coupling theory, Pitts {\em et al.}
have presented a diagrammatic treatment that yields similar equations
to those of mode coupling theories of the glass
transition\cite{PittsAndersen01}.  This treatment lends itself to a
closure assumption that is very similar kind to the mode coupling
theory of G\"otze {\em et al}.  Pitts {\em et al.} propose alternative
closure assumptions to this mode coupling theory by summing subsets of
diagrams. The resulting predictions are in good agreement with
simulations for high concentration of up-spins, but still decay too
rapidly at lower spin density. Recent improvements on this scheme have
been carried out by Wu and Cao\cite{WuCao04} based on a combination of
matrix methods and mode coupling closure assumptions.

Regardless of the precise formalism, mode coupling theories aim to
describe slow long time behavior, and so should include all the ``slow
modes'' of the system. For hydrodynamics of a simple fluid at moderate
densities, these slow modes are well-known, namely the density,
momentum and energy modes at large wave lengths. Correspondingly, when
applied to Fourier modes of these hydrodynamical fields, mode coupling
theory (e.g. in the formulation of Oppenheim and co-workers)
yields well defined perturbative results when the correlation length
is finite and the thermodynamic limit is taken.  However, in the case
of the East model, the relevant slow modes are less obvious, and
previously developed mode coupling theories for this
model\cite{JaeckleEisinger91,Kawasaki95,Pittsetal00,PittsAndersen01,WuCao04}
may have missed some of these slow modes. In fact the absence of some
slow modes provides a possible explanation for the problem these
theories have with describing the long time behavior at small $c$.

The main purpose of this article is to identify the relevant slow
modes in frustrated spin systems and to describe the impact of the
coupling of these modes to a specific spin variable.  It is
demonstrated that the existence of slowly-relaxing spin domains of
arbitrary size suggests a natural basis of slow modes in which
quantitatively accurate but simple approximation schemes are easily
formulated for many quantities of interest in the study of slow
heterogeneous relaxation.

\section{\label{eastmodel}The East model}

The East model\cite{JaeckleEisinger91} is a linear chain of $N$ spins,
which are numbered from $0$ to $N-1$, with each spin allowed to assume
one of two values at any given time, here taken to be up or
down. Occupation numbers $n_i$ are defined such that $n_i=1$ when spin
$i$ is up and $0$ if it is down.  The static properties of this model
follow from the Hamiltonian
\begin{equation}
  H =  \mu \sum_{i=0}^{N-1} n_i.
\label{Hamiltonian}
\end{equation}
Using the canonical distribution $\rho \sim \exp[-\beta H]$, the
average occupation per site if the system is at equilibrium at an
inverse temperature $\beta$, is found to be
\begin{equation}
  c = 1/(1+e^{\beta\mu}).
\label{cbm}
\end{equation}
As $\beta\mu$ has little physical significance in the current context,
the density $c$ will be used as a parameter.  If $\mathbf n$ denotes a
spin state (i.e., a configuration of the $N$ spins), the
canonical equilibrium distribution can be written as
\begin{equation}
  \rho(\textbf n) =
  \prod_{i=0}^{N-1} \left[c\, n_i + (1-c)\,(1-n_i)\right]
\label{equilibrium}
\end{equation}
where Eqs.~\eqref{Hamiltonian} and \eqref{cbm} were used, as well as
the fact that $n_i$ is either zero or one. Note that each spin $i$ has
a probability $c$ to be up ($n_i=1$) and $1-c$ to be down ($n_i=0$).

For the dynamics, consider the conditional probability density
$U_t(\mathbf n,\mathbf n')$ to be in state $\mathbf n$ at time $t$
given that the system was in state $\mathbf n'$ at time $0$, which
satisfies \cite{PittsAndersen01,VanKampen}
\begin{eqnarray}
  \dot U_t(\mathbf n,\mathbf n')
  &=& \sum_{\tilde{\mathbf n}}\Big[
   W(\mathbf n,\tilde{\mathbf n})U_t(\tilde{\mathbf n},\mathbf n')
    -W(\tilde{\mathbf n},\mathbf n)U_t(\mathbf n,\mathbf n')\Big]
\nonumber
\\
  &\equiv&
  \sum_{\tilde{\mathbf n}} \mathcal L(\tilde{\mathbf n},\mathbf n)
  U_t(\tilde{\mathbf n},\mathbf n'),
\label{master}
\end{eqnarray}
with $U_0(\mathbf n,\mathbf n')=\delta_{\mathbf n\mathbf n'}$.  Here,
$W(\mathbf n,\mathbf n')dt$ is the probability to make a transition
from $\mathbf n'$ to $\mathbf n$ in a time $dt$. By definition,
$W(\mathbf n,\mathbf n)=0$.  Defining
\begin{equation}
        A(\mathbf n,t) = \sum_{\mathbf n'}
        A(\mathbf n') U_t(\mathbf n',\mathbf n),
\label{Aev}
\end{equation}
from Eq.~\eqref{master} it follows that
\begin{equation}
        \dot A(\mathbf n,t) = \sum_{\mathbf n'}
\mathcal L(\mathbf n,\mathbf n') A(\mathbf n',t)        .
\nonumber
\end{equation}
or
\begin{equation}
\dot A(t)=\mathcal LA(t)
\label{ddtALA}
\end{equation}
where the Liouville operator $\mathcal L$ is a linear operator on the
$2^N$ dimensional Hilbert space of functions of $\mathbf n$.  The
formal solution of Eq.~\eqref{ddtALA} is $A(t)=e^{\mathcal Lt}A(0)$.

For the East model, $W(\mathbf n,\mathbf n')$ can be written as a sum
over possible moves, i.e., spin flips of individual sites $i$:
\begin{equation}
        W(\mathbf n,\mathbf n')
        = \sum_{i=0}^{N-1} W_i(\mathbf n,\mathbf n').  
\nonumber
\end{equation}
A flip of spin $n_i$ is possible only if $n_{i+1}=1$ (using the
boundary condition that $n_{N}=1$), as the expression for $W_i$ bears
out\cite{PittsAndersen01}:
\[
        W_i(\mathbf n,\mathbf n') = \left[ c \delta_{n_i1}\delta_{n_i'0} +
        (1-c)\delta_{n_i0}\delta_{n_i'1}\right] n_{i+1}
        \prod_{j\neq i} \delta_{n_j n_j'}.
\]
Correspondingly, the Liouville operator can be written as
\begin{eqnarray}
        \mathcal L(\mathbf n,\mathbf n')
        &=& \sum_{i=0}^{N-1}
        \Big[
        (1- c) \delta_{n_i1}\left(\delta_{n_i'0}-\delta_{n_i'n_i}\right)
\label{Ldef}
\\&&
+c \delta_{n_i0}\left(\delta_{n_i'1}-\delta_{n_i'n_i}\right)
        \Big]
n_{i+1}         \prod_{j\neq i} \delta_{n_j n_j'}.
\nonumber
\end{eqnarray}

The equilibrium distribution in Eq.~\eqref{equilibrium} will serve as
a weight for the inner product on the Hilbert space, i.e., the
inner product of $A(\mathbf n)$ and $B(\mathbf n)$ is $\inner{A}{B} =
\sum_{\mathbf n} \rho(\mathbf n) A(\mathbf n) B(\mathbf
n)\equiv\average{A B}$. Only real quantities will be used in this
paper, so there is no need to define a complex inner product (although
this is straightforward). The time correlation function of $A$ and $B$
can now be written as $\inner{A(t)}{B}$.  When $W(\mathbf n,\mathbf
n')\rho(\mathbf n') = W(\mathbf n',\mathbf n)\rho(\mathbf n)$, {\it
i.e.}, when detailed balance holds, as it does in this model,
$\mathcal L$ is Hermitian with respect to the inner product. In
contrast to stochastic systems, in deterministic systems the Liouville
operator is anti-Hermitian.  It should be noted that the condition of
detailed balance also guarantees that the limiting stationary
distribution of the Markovian dynamics is the equilibrium
distribution~\eqref{equilibrium}, provided the underlying Markov
process is ergodic\cite{VanKampen}.

As remarked by Pitts et al.\cite{Pittsetal00}, different sites are not
only \emph{statically} independent [cf. Eq.~\eqref{Hamiltonian}], but
also \emph{dynamically}. To see this, consider the normalized single
site fluctuation
\begin{equation}
        \hat n_{i}(t) = \frac{n_i(t)-c}{\sqrt{c(1-c)}}.
\label{fluct}
\end{equation}
which satisfies $\average{\hat n_{i}^2(t)}$~$=$~$1$
and $\average{\hat n_{i}(t)}=0$.  
Note that
\begin{eqnarray}
  \mathcal L \hat n_{i} &=&   -n_{i+1}\hat n_{i}
\label{second}
\\
        &=& -c \hat n_{i} + \sqrt{c(1-c)}\,\hat n_{i}\hat n_{i+1},
\label{esta}
\end{eqnarray}
using Eqs.~\eqref{Ldef} and \eqref{fluct}. Thus, the time derivative
of $\hat n_{i}$ depends on the product of $\hat n_{i}$ and $\hat
n_{i+1}$. The derivative of that product will in turn depend on $\hat
n_{i+2}$, and so on. Thus, $\hat n_{i'}(t)=e^{\mathcal Lt} \hat
n_{i'}$ involves only $\hat n_{i}$ with $i\geq i'$. From the static
independence of any site $i$ and any other site $i''$ with $i''<i'\leq
i$, it follows that $\average{\hat n_{i''}\hat
n_{i'}(t)}=\average{\hat n_{i''}}\average{\hat n_{i'}(t)}=0$.  Because
$\mathcal L$ is Hermitian, also $\average{\hat n_{i''}(t)\hat
n_{i'}}=0$.  So all time correlation functions between different sites
$i'\neq i''$ are zero.

Given the dynamical independence of different sites, we are interested
in the nontrivial time correlation function
\begin{equation}
        C(t) = \average{\hat n_{i}(t)\hat n_{i}(0)}
\label{Ctdef}.
\end{equation}
In the limit $N\to\infty$ with $i$ fixed, this is independent of $i$
due to translation invariance. This single spin time correlation
function in the thermodynamic limit is the main quantity of interest.

\section{\label{modecoupling}Projection Operator Techniques}

In a mode coupling framework, dynamical equations are derived for the
time correlation functions of slow modes in the system, which involves
a memory kernel that is again expressed in terms of the correlation
functions\cite{Keyes77,Zwanzigbook,Pittsetal00}.  The starting point
is often a projection operator formalism. Here, a general setup will
be presented, which uses the Mori-Zwanzig projection operator
formalism\cite{Mori,Zwanzigbook}, to be specialized later.

Let $A_k$ be the slow modes of the system determined from physical
arguments, with $k$ an index running over the slow modes.  It will be
assumed that $\average{A_k}=0$ (as could be achieved by subtracting
the average), and that $A_k$ are orthonormal, i.e.,
$\inner{A_k}{A_q}=\delta_{kq}$ (as could be achieved by a
Gramm-Schmidt procedure).  For brevity, the $A_k$'s are taken together
in a vector $A$.  In the projection operator formalism, the component
along $A$ of any other physical quantity $B$ is found using the
projection operator
\begin{equation}
  {\mathcal P} B = \inner{B}{A}\cdot A = \sum_k\inner{B}{A_k} A_k,
\label{P1def}
\end{equation}
where $\cdot$ denotes a vector product, i.e., a sum over $k$, as
indicated.  Using the operator identity
\begin{equation}
   e^{\mathcal L t} = e^{(1-\mathcal P)\mathcal L t} + \int_0^t
        e^{\mathcal L \tau}
        \mathcal{P} L
        e^{(1-\mathcal P)\mathcal L (t-\tau)}
        d \tau,
\label{operatoridentity}
\end{equation}
and Eqs.~\eqref{ddtALA} and \eqref{P1def}, one can derive that
\begin{equation}
        \dot A(t) = \mathsf M^E \cdot A(t) + \int_0^t
        \mathsf M^D(t-\tau)\cdot
        A(\tau) d\tau + \varphi(t),
\label{memorydiff}
\end{equation}
where $\varphi(t) = e^{(1-\mathcal P)\mathcal L t}(1-\mathcal
P)\mathcal L A$ and
\begin{eqnarray}
   \mathsf M^E &=& \mel{A}{\mathcal L}{A}
\label{ME11}
\\
   \mathsf M^D(t) &=& \inner{\varphi(t)}{\varphi}
\label{MD11}.
\end{eqnarray}
Note that $\mathsf M^E$ and $\mathsf M^D(t)$ are matrices whose dimensions
are equal to the number of slow modes and that $\mathsf M^E$ contains
only static information, while the memory kernel $\mathsf M^D(t)$
involves the time correlation of the fluctuating force $\varphi(t)$.

Taking the inner product with $A$, Eq.~\eqref{memorydiff} yields an
equation for the correlation function $\mathsf
G(t)\equiv\inner{A(t)}{A}$:
\begin{equation}
   \dot{\mathsf G}(t) = \mathsf M^E\cdot\mathsf G(t)
        + \int_0^t \mathsf M^D(t-\tau)\cdot \mathsf G(\tau) d\tau.
\label{memorycorr}
\end{equation}
To solve this equation, a Laplace transform will be used, defined as
\begin{equation}
\tilde{\mathsf G}(z) = \int_0^\infty e^{-zt} \mathsf G(t) \,dt.
\label{laplace}
\end{equation}
Here and the following, we adopt the convention that quantities with a
tilde ($\tilde{\ }$) are $z$-dependent.
The solution in Laplace space of Eq.~\eqref{memorycorr} is
\begin{equation}
  \tilde{\mathsf G} =
    (z\identity-\mathsf M^E-\tilde{\mathsf M}^D)^{-1}.
\label{laplacesol}
\end{equation}
where
\begin{equation}
  \tilde{\mathsf M}^D = \int_0^\infty e^{-zt} \mathsf M^D(t) \,dt.
\end{equation}

\section{Physics of the Slow Dynamics}

If $A$ in the previous section contained all the slow behavior, then
$\mathsf M^D(t)$ would be a quickly-decaying function that could be
replaced by a delta function in time and integrated over in
Eq.~\eqref{memorycorr}\cite{Keyes77,Zwanzigbook}. Unfortunately, this
is typically not the case because the projection operator $(1-\mathcal
P)$ only removes part of the dependence on $A$.  For instance in a
fluid, the long wave length modes of density, momentum and energy are
slow because they correspond to densities of conserved quantities, but
only at low densities is it enough to consider only these modes as
slow. Extending the set $A$ by multi-linear
modes\cite{MachtaOppenheim82,KavassalisOppenheim88}, can help, and can
be used to setup self-consistent equations which are exact in the
thermodynamic limit provided there is a finite dynamical correlation
length\cite{MachtaOppenheim82,Schofieldetal92,LiuOppenheim97,LiuOppenheim97b}.

As is known from the extensive work of G\"otze and co-workers
\cite{GotzeLucke75,Bosseetal78a,Bengtzeliusetal84,Leutheusser84,GotzeSjoegren87,GoetzeSjoegren88,Fuchsetal91,GoetzeSjoegren92,Cumminsetal93,Fuchsetal98,Goetze99,GoetzeSperl02}
in deterministic systems, such self-consistent equations can give rise
to a glass transition.

If, on the other hand, $A$ is a complete set, then $1-\mathcal P=0$
and consequently $\varphi(t)=0$ and $\mathsf M^D(t)=0$.  In this case,
the above formalism corresponds to writing Eq.~\eqref{ddtALA} in a
particular basis. This formulation is often applied to the East
model\cite{Pittsetal00,PittsAndersen01,WuCao04}. When working with a
complete basis set, the set still has to be truncated at some level in
practice. This introduces truncations errors, or, viewed
alternatively, a nonzero $\mathsf M^D(t)$. To get beyond the truncation
problem, one makes an ansatz for the memory kernel in terms of the
time correlation function of interest [here $C(t)$], yielding a
self-consistent equation.  However, in stochastic systems, a glass
transition will not be found if such an ansatz is used for the memory
kernel, due to the Hermitian nature of $\mathcal
L$\cite{WuCao04,PittsAndersen01}. Rather, an ansatz needs to be used
for the so-called ``irreducible'' memory
kernel\cite{Kawasaki95,PittsAndersen00}. Then, a glass transition can
be found for finite $c$ in the East
model\cite{Pittsetal00,PittsAndersen01}. However, simulations make it
clear that there is no transition to a non-ergodic phase at a non-zero
value of $c$. Somewhat better schemes to improve the ansatz have been
developed since\cite{WuCao04}, but generally, they lead either to a
transition or to a time correlation function that decays too quickly.

Essential for the success of mode coupling theories for fluids at
lower densities is the finiteness of the dynamical correlation length,
which gives a cutoff length and certain exact factorization
properties in the thermodynamic
limit\cite{MachtaOppenheim82,Schofieldetal92}.  However, for the East
model, no such length scale exists, as is seen when one contrasts the
result of the scaling of the decay time
\cite{SollichEvans99,AldousDiaconis02} vs. the diagrammatic approach
of Pitts and Anderson\cite{PittsAndersen01}. When diagrams are
truncated at a certain level, corresponding to taking into account
only spins within a certain distance $l$, the memory kernel becomes a
polynomial in $c$ of which highest power is $c^l$, which means the
typical time can scale at the slowest as $c^{-l}$. However, for
$c\to0$, the timescale diverges faster than any inverse power of
$c$\cite{SollichEvans99,AldousDiaconis02}. So clearly the spins at all
positions, arbitrarily far away, need to be taken into account.

A second way to see that the dynamic correlation length is unbounded
is to note that while the static correlation length is zero,
dynamically, the decay of the time correlation function $C(t)$
\emph{is} influenced by other spins. For example, if there is a large
domain of down spins to the east of a given spin, that particular spin
requires a long time to flip since all down spins in the domain must
flip at least once.  Hence, the decay is correlated with the existence
of this domain, and the dynamic correlation length is therefore at
least of the order of the size of this domain. But domains of all
sizes exist and larger sized domains will contribute to the behavior
of the time correlation function $C(t)$ at longer times.  Even if one
is only interested in the bulk of the behavior of the time correlation
function $C(t)$, for which the relevant domains are of typical size
$1/c$, this size diverges as $c\to0$. Therefore, it is no surprise
that fixed spatial truncations do not work below a certain value of
$c$ and that mode coupling theories using such truncations have
problems to describe the long time behavior of the time correlation
function $C(t)$.  For a different formulation of the importance of
domains, see Garrahan and
Chandler\cite{GarrahanChandler02,GarrahanChandler03b} and Wu and
Cao\cite{WuCao04}.

Thus, physically, the origin of the slowness of the dynamics seems to
be related to the absence of a finite dynamical correlation length and
the existence of arbitrarily large domains of down spins.

\section{\label{gapbasis}The domain Basis}

\subsection{\label{singlegap}Single domains}

Consider the leftmost spin $n_0$ in a semi-infinite chain of spins.
East of this leftmost spin (i.e., at sites $i > 0$), a domain of
typical size $1/c$ filled with down spins exists.  In the previous
section it was argued that the presence of these domains is essential
to the dynamics, so they should somehow be included. This is achieved
by defining the (single) {\em domain basis}, which is composed of the
orthonormal basis vectors
\begin{eqnarray}
        \hat Q_0 &=& \frac{1}{Z_0}(n_0-c)
        \label{A0def}
\end{eqnarray}
and
\begin{subequations}
\label{20}
\begin{eqnarray}
        \hat Q_1(0) &=& \frac{1}{Z_1(0)}(n_0 - c) (n_1 - c)\\
        \hat Q_1(1) &=& \frac{1}{Z_1(1)}(n_0 - c) (1-n_1)(n_2-c)
\end{eqnarray}
and in general
\begin{equation}
        \hat Q_1(k) =  \frac{1}{Z_1(k)}(n_0 - c)
        \prod_{j=1}^{k}(1-n_j)
                (n_{k+1}-c).
 \label{Akdef}
\end{equation}        
\end{subequations}
where the normalization constants are to be chosen such that
$\average{\hat Q_0^2}=\average{[\hat Q_1(k)]^2}=1$.  Note that in
Eqs.~\eqref{20} each factor $(1-n_j)$ only yields a contribution when
$n_j=0$, \emph{i.e.}, when spin $j$ is down. Thus, a consecutive
sequence of such factors represents a down-spin domain.

Comparing the basis vector in Eq.~\eqref{A0def} to $\hat n_{0}$ in
Eq.~\eqref{fluct}, which is normalized, it is seen that $\hat Q_0$
should be equal to $\hat n_{0}$, i.e.,
\begin{equation}
  Z_0 = \sqrt{c(1-c)}\label{z0def}.
\end{equation}
Using a convenient diagrammatic representation we will later derive
that
\begin{equation}
  Z_1(k) = c(1-c)^{1+k/2}.  \label{z1def}
\end{equation}
Other useful quantities will be the unnormalized versions of these
basis vectors
\begin{eqnarray}
  Q_0 &=& (n_0-c)
\label{unz0}
\\
  Q_1(k) &=&  (n_0 - c)        \prod_{j=1}^{k}(1-n_j)
                (n_{k+1}-c).
 \label{unz1}
\end{eqnarray}
Note that
\begin{subequations}
\label{starstar}
\begin{eqnarray}
   \hat Q_0&=&Q_0/Z_0\\
   \hat Q_1(k)&=&Q_1(k)/Z_1(k).
\label{stars}
\end{eqnarray}
\end{subequations}

It is convenient at this point to introduce a diagrammatic notation
for the unnormalized quantities $Q_0$ and $Q_1$, depicted in
Table~\ref{diagramE}.  In such diagrams, the horizontal direction
represents the lattice.  An open circle ($\circ$) at a certain
position $i$ denotes $(n_i-c)$, a horizontal line
($\textdline{2}{\cbar}$) denotes $(1-n_i)$ and a product of
consecutive $(1-n_i)$'s is represented as a longer horizontal line
with the number of factors written on top of the line.  When these
diagrammatic elements are touching, this denotes that they are on
neighboring lattice sites. The first (i.e., leftmost) symbol in
a diagram will always refer to site~$0$. Note that after a diagram for
$Q_0$ and $Q_1$ has been evaluated, one still has to divide by the
appropriate normalization to get the quantities corresponding to $\hat
Q_0$ and $\hat Q_1$, cf.~Eqs.~\eqref{starstar}.

\begin{table}[tb]
\hrule
\begin{tabular}{c@{\ \ \ }@{\ \ \ }c@{\ \ \ }@{\ \ \ }c@{\ \ \ }@{\ \ \ }c}
\dline{2}{\setO}    &
\dline{3}{\setO\setO} &
\dline{4}{\setO\cbar\setO} &
\dline{7}{\setO\cbar\txt{k}\cbar\setO}
\\
$Q_0$
& $Q_1(0)$
& $Q_1(1)$
& $Q_1(k)$
\end{tabular}
\hrule
\caption{Diagrammatic representation of the domain basis.}\label{diagramE}
\end{table}        

As an example, this diagrammatic notation will be used to show that
Eq.~\eqref{z1def} is the correct choice for $Z_1(k)$ such that
$\inner{\hat Q_1(k')}{\hat Q_1(k)}$ is equal to $\delta_{kk'}$.  The
product of $Q_1(k)$ and $Q_1(k')$ will be represented as the two
diagrams of $Q_1(k)$ and $Q_1(k')$ from Table~\ref{diagramE} on top of
each other:
\begin{equation}
 \inner{Q_1(k')}{Q_1(k)}
  =
  \inner{\dline{6}{\setO\cbar\txt{k'}\cbar\setO}
  }{ \dline{7}{\setO\cbar\cbar\txt{k}\cbar\cbar\setO}}
  =
    \diagram{7}{
        \setO\cbar\cbar\txt{k}\cbar\cbar\setO\nxt
        \setO\cbar\ltxt{k'}\cbar\setO
    }
\label{Kdiagram}
\end{equation}
The vertical displacement of the two parts in the diagram on the right
hand side serves to distinguish one part from the other and to
indicate that the diagram needs to be averaged.  Next, note that since
$\average{\circ}=\average{n_i-c}=0$, any open circle at site $i$ in
the diagram has to overlap with an open circle or a horizontal line at
the same site $i$ from the other part in order to give a non-vanishing
contribution to the average. In the diagram in Eq.~\eqref{Kdiagram},
this can only happen if $k=k'$. Thus, the domain basis is
orthogonal. The case $k=k'$ is represented by
\begin{equation}
 \inner{Q_1(k)}{Q_1(k)}
   =
        \diagram{8}{
         \setO\cbar\cbar\txt{k}\cbar\cbar\setO \nxt
         \setO\cbar\cbar\cbar  \cbar\cbar\setO
        }
   = c^2(1-c)^{k+2}.
\label{28prime}
\end{equation}
In the last equality, we used that different parts of a diagram have
different contributions depending on which diagrammatic elements on
top align with which elements on the bottom, as shown in
Table~\ref{rules} (the diagrammatic element ``$\bullet$'' will be
introduced later).  The contributions of the various parts are simply
multiplied because each site is (statically) independent.  From
Eqs.~\eqref{stars} and \eqref{28prime} it follows that
\begin{equation}
 \inner{\hat Q_1(k)}{\hat Q_1(k)} =
   \frac{\inner{Q_1(k)}{Q_1(k)}}{[Z_1(k)]^2} =
  \frac{c^2(1-c)^{k+2}}{[Z_1(k)]^2}.
\end{equation}
With the definition of $Z_1(k)$, Eq.~\eqref{z1def}, we immediately
see that the $\hat Q_1(k)$ are properly normalized.
In a similar way, one can deduce that $\inner{\hat
Q_0}{\hat Q_1(k)}=0$, thus establishing that the basis is really
orthonormal.

\begin{table*}[t]
\begin{tabular}{l|c@{\ \ \ }c@{\ \ \ }c@{\ \ \ }c@{\ \ \ }c@{\ \ \ }c}
\hline
element:
&\textdline{2}{\setO}
&\textdline{2}{\cbar}
&\textdline{2}{\setX}
&(nothing)
&
&
\\
meaning:
&$n-c$
&$1-n$
&$n$
&1
&
&
\\
when averaged, results in:
&$0$
&$1- c$
&$c$
&1
&
\\\hline
horizontal part:
&\diagram{2}{\cbar\nxt\cbar}
&\diagram{2}{\setX\nxt\setO}
&\diagram{2}{\setX\nxt\cbar}
&\diagram{2}{\setX\nxt\setX}
&\diagram{2}{\setO\nxt\setO}
&\diagram{2}{\setO\nxt\cbar}
\\
meaning:
&$(1-n)^2$
&$n(n-c)$
&$n(1-n)$
&$n^2$
&$(n-c)^2$
&$(1-n)(n-c)$
\\
when averaged, results in:
&$ (1-c)$
&$ c(1-c)$
&$ 0$
&$c$
&$ c(1-c)$
&$-c(1-c)$\\
\hline
\end{tabular}
\caption{Rules for the contributions to a diagram. On top are the
  elementary diagrammatic elements, at the bottom combinations of
  them.  The resulting factors are obtained using
  Eq.~\eqref{equilibrium}.  } \label{rules}
\end{table*}

Taking the collection $\{\hat Q_0, \hat Q_1\}$ for $A$ in the
projection formalism of Sec.~\ref{modecoupling}, the matrix $\mathsf
M^E$ in Eq.~\eqref{ME11} which determines the dynamics, becomes
\begin{equation}
 \mathsf M^E
= \left[\begin{matrix}
                \mel{\hat Q_0}{\mathcal L}{\hat Q_0} &
                \mel{\hat Q_0}{\mathcal L}{\hat Q_1} \\
                \mel{\hat Q_1}{\mathcal L}{\hat Q_0} &
                \mel{\hat Q_1}{\mathcal L}{\hat Q_1}
                \end{matrix}\right],
\label{thiscase}
\end{equation}
where $\hat Q_1$ without a value of $k$ denotes the column vector
(in the ket) or row vector (in the bra) composed of all $\hat
Q_1(k)$.  From Eqs.~\eqref{Ldef} and \eqref{second} it follows that
\begin{subequations}
\label{27}
\begin{eqnarray}
  \mathcal L Q_0 &=&- (n_0-c) n_1
\label{LQ0}
=- \,\dline{3}{\setO\setX}
\\
        \mathcal L Q_1(0) &=& - (n_0-c) [(n_1-c) n_{2}
                                         + n_{1}  (1-c)]
\nonumber\\
&=&- \,\dline{4}{\setO\setO\setX} - (1-c) \,\dline{2}{\setO\setX}
\label{LQ10}
\\
        \mathcal L Q_1(k\geq1)  &=&  (n_0-c)
                           (1-n_1)\cdots(1-n_{k-1})
\nonumber\\&&\quad\times
        [ -(1-n_{k}) (n_{k+1}-c) n_{k+2}
\nonumber\\&&\qquad
          +  (n_{k}-c) n_{k+1} (1-c)]
\nonumber\\
&=&
 -   \dline{10}{\setO\cbar\cbar\txt{k}\cbar\cbar\setO\setX}
            + (1-c)
                \dline{8}{\setO\txt{k-1}\cbar\cbar\cbar\cbar\setO\setX}
\label{LQ1k}
\end{eqnarray}
\end{subequations}
In the diagrammatic representation in Eqs.~\eqref{LQ0}--\eqref{LQ1k},
a solid circle $\bullet$ denotes $n_{k+1}$ or $n_{k+2}$. 

A few words are in order on how to obtain the diagram of $\mathcal L
X$ given that of $X$. The Liouville operator $\mathcal L$ acts much
like a differential operator and can be shown to follow the product
rule $\mathcal L (AB)=A(\mathcal L B)+(\mathcal L A)B$, provided $A$
and $B$ do not involve the same site.  Thus, $\mathcal L$ acting on
diagrams like those in Table~\ref{diagramE} yields a sum of terms
where $\mathcal L$ acts on each site individually. Each site has one
of the diagrammatic elements $\circ$, $\bullet$ or
$\textdline{2}{\cbar}$, and
\begin{equation}
  \mathcal L \dline{2}{\setO} =  - \dline{3}{\setO\setX}, \quad
  \mathcal L \dline{2}{\setX} =  - \dline{3}{\setO\setX},\quad
  \mathcal L \dline{2}{\cbar} =  \dline{3}{\setO\setX}.
\end{equation}
Thus, $\mathcal L$ acting on site $i$ introduces a new diagrammatic
element $\bullet$ in the diagram on the next site $i+1$. If the diagram
already had an element on that site, one needs to multiply the new one
with the original one, and this gives the three possibilities
\begin{equation}
  \dline{2}\setX \times \dline{2}\setO  =  (1-c)\dline{2}\setX, \quad
  \dline{2}\setX \times \dline{2}\setX =  \dline{2}\setX,\quad
  \dline{2}\setX \times \dline{2}\cbar =  0.
\label{mults}
\end{equation}
As the last equation shows, $\mathcal L$ acting on an element at site
$i$ yields zero if site $i+1$ has the diagrammatic element
``$\textdline{2}{\cbar}$''. This is the reason why there are so few
diagrams in Eqs.~\eqref{LQ1k}: although $\mathcal L$ could in
principle act on all sites in $Q_1(k)$, most have an element
``$\textdline{2}{\cbar}$'' to their right and yield zero.

The matrix elements $\mel{Q_0}{\mathcal L}{Q_0}$, $\mel{Q_0}{\mathcal
L}{Q_1(k)}$ and $\mel{Q_1(k')}{\mathcal L} {Q_1(k)}$ are found by
taking the inner product of the diagrams in Eq.~\eqref{27} with those
of $Q_0$ and $Q_1(k')$ in Table~\ref{diagramE}. Because each open
circle needs to be covered, the only nonzero contributions are for
$k'=k-1, k $ and $k+1$, and
\begin{subequations}\label{ALA}
\begin{eqnarray}
\mel{Q_0}{\mathcal L}{Q_0} &=&
        -\diagram{3}{
          \setO\setX    \nxt
          \setO
        }
=-c^2(1-c)
\\
        \mel{Q_1(0)}{\mathcal L}{Q_0} &=&
        - \diagram{3}{
          \setO\setX \nxt
          \setO\setO
          }
=-c^2(1-c)^2
\\
        \mel{Q_1(0)}{\mathcal L}{Q_1(0)} &=&
        -\diagram{4}{
          \setO\setO\setX \nxt
          \setO\setO
        }
        - (1-c)
        \diagram{3}{
        \setO\setX \nxt
        \setO\setO
        }
\nonumber\\&=&
- c^2(1-c)^2
\\
\!        \mel{Q_1(k)}{\mathcal L}{Q_1(k)} &=&
        -\diagram{8}{
        \setO\cbar\cbar\txt{k}\cbar\cbar\setO\setX \nxt
        \setO\cbar\cbar\cbar\cbar\cbar\setO
        }
        +(1-c)\diagram{7}{
        \setO\txt{k-1}\cbar\cbar\cbar\cbar\setO\setX \nxt
        \setO\cbar\cbar\cbar\cbar\cbar\cbar\setO
        }
\nonumber\\&=&
-c^3(2-c)(1-c)^{k+2}\text{ if }k>0
\end{eqnarray}
and
\begin{equation}
        \mel{Q_1(k+1)}{\mathcal L}{Q_1(k)} =
        -\diagram{9}{
        \setO\cbar\cbar\txt{k}\cbar\cbar\setO\setX \nxt
        \setO\cbar\cbar\cbar\cbar\cbar\cbar\setO
        }
=
  c^3(1-c)^{k+3},
\end{equation}
\end{subequations}
where the rules in Table~\ref{rules} were used.  Using
Eqs.~\eqref{starstar}, \eqref{thiscase} and \eqref{ALA}, $\mathsf M^E$
becomes the infinite tridiagonal matrix
\begin{equation}
\! \mathsf M^E =
 \left[\begin{matrix}
 - c            &-\sqrt{c(1-c)}\\
 -\sqrt{c(1-c)}&-1            &c\sqrt{1-c}\\
               &c\sqrt{1-c}   &-c(2-c)    &\ddots\\
               &              &\ddots     &\ddots
 \end{matrix}\right],
\label{M11Egap}
\end{equation}
where diagonal dots denote repetition of the last mentioned expression
on that diagonal.

At this stage, $\mathsf M^D(t)$ in Eq.~\eqref{memorycorr} and
$\tilde{\mathsf M}^D$ in Eq.~\eqref{laplacesol} will be set to zero. The
reasons for this are twofold. First, it allows an exact solution for
spin autocorrelation functions to be obtained that is in good
quantitative agreement with simulations if the density of up-spins $c$
is not too low. Secondly, we will later complete the basis such that
$\tilde{\mathsf M}^D$ is in fact strictly zero. Eq.~\eqref{laplacesol}
yields in this approximation:
\begin{equation}
   \tilde{\mathsf G} \approx    \tilde{\mathsf G}^{(1)}
\equiv (z\identity-\mathsf M^E)^{-1}.
\label{Glaplace}
\end{equation}
Here, the superscript $(1)$ indicates that the result is only a first
approximation. Below we will make this into a systematic approximation
scheme in which further, more accurate approximations can be obtained.

According to Eq.~\eqref{A0def}, the Laplace transform of the
correlation $C(t)$ in Eq.~\eqref{Ctdef} is the top left element of the
matrix $\tilde{\mathsf G}$,
\begin{equation}
\tilde C = \int_0^\infty dt\, e^{-zt} C(t) = \tilde{\mathsf G}_{11}.
\end{equation}

\begin{figure}[b]
\centerline{\includegraphics[angle=-90,width=0.43\textwidth]{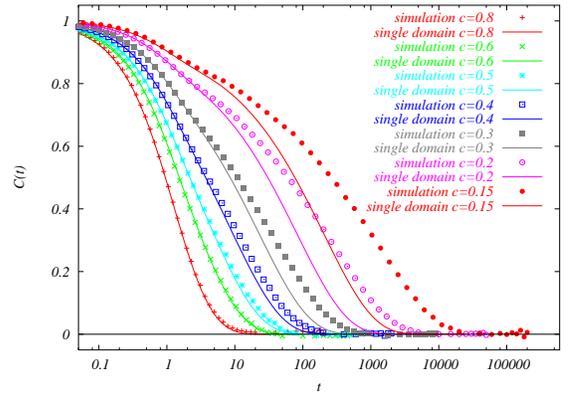}}
\caption{Results for the single spin time correlation function $C(t)$
from the single domain basis $\{\hat Q_0,\hat Q_1\}$ [by numerical
Laplace inversion of Eq.~\eqref{Cz} in Sec.~\ref{singlegap} using {\it
Mathematica}], compared to simulation data (kindly provided by
Prof. H.~C.~Anderson).}
\label{Mathematicaresults}
\end{figure}

To perform the matrix inversion in Eq.~\eqref{Glaplace}, one uses the
fact that the inverse of a tridiagonal matrix can be performed
exactly. In particular, the top left element of the inverse of a
symmetric tridiagonal matrix can be written as a continued fraction:
\begin{equation}
\left[\begin{matrix}
a_1&b_1\\
b_1&a_2&b_2\\
   &b_2&a_3&\ddots\\
   &   &\ddots&\ddots
\end{matrix}\right]
^{-1}_{11} =
\cfrac{1}{a_1 - \cfrac{b_1^2}{a_2 - \cfrac{b_2^2}{a_3 - \ldots}}}.
\label{contfrac}
\end{equation}
Combining Eq.~\eqref{M11Egap}--\eqref{contfrac}, one finds
\begin{eqnarray}
  \tilde C^{(1)} &=&
  \left[\begin{matrix}
      z+c         & \sqrt{c(1-c)}\\
      \sqrt{c(1-c)}&  z+1         &- c\sqrt{1-c}\\
             &- c\sqrt{1-c}  & z+c(2-c) & \ddots\\
             &              &\ddots    &\ddots
    \end{matrix}\right]
  ^{-1}_{11}
\nonumber\\
  &=&\cfrac{1}{z+c -\cfrac{c(1-c)}{z+1-\cfrac{c^2(1-c)}{z+c(2-c)
      -\frac{c^2(1-c)}{z+c(2-c)-\ldots}}}}
\label{Cprel}
\end{eqnarray}
The repeating part of this expression is
\begin{equation}
 \tilde\gamma^{(1)} = \cfrac{c^2(1-c)}{z+c(2-c)-\cfrac{c^2(1-c)}{z+c(2-c)-\ldots}}.
\label{gammadef}
\end{equation}
This $\tilde\gamma^{(1)}$ clearly satisfies
\begin{equation}
  \tilde\gamma^{(1)} = \frac{c^2(1-c)}{z+c(2-c)-\tilde\gamma^{(1)}}
\label{gammaeq}
\end{equation}
which is solved by
\begin{equation}
  \tilde\gamma^{(1)} = \frac12\left\{z+c(2-c)-\sqrt{4cz+(z - c^2)^2}\right\}.
\label{gammasol}
\end{equation}
[Note that the solution of Eq.~\eqref{gammaeq} with a plus sign in
front of the square root in Eq.~\eqref{gammasol} does not go as $1/z$
for large $z$, and is therefore not in agreement with 
Eq.~\eqref{gammadef}].  Inserting this result in Eq.~\eqref{Cprel},
one obtains the explicit form
\begin{equation}
  \tilde C^{(1)} =  \cfrac{1}
{z+c - \cfrac{2c(1-c)}{z+2-(2-c)c+\sqrt{4cz+(z-c^2)^2} }}
\label{Cz}
\end{equation}

The exact correlation function in Eq.~\eqref{Cz} can be Laplace
inverted numerically using Stehfest's
algorithm\cite{Stehfest70a,Stehfest70b} (for example, in
\emph{Mathematica}\cite{StehfestMathematicaPackage}).  For various
values of $c$, the results are shown in Fig.~\ref{Mathematicaresults}
and compared with data from simulations on the East model. Despite the
simple form of $\tilde C^{(1)}$ in Eq.~\eqref{Cz}, there is excellent
agreement between this theoretical result and the data for $0.6\leq c
\leq 1$, reasonable qualitative agreement up to $c\approx 0.5$, while
the predicted decay is clearly too fast for $c<0.5$.

\subsection{\label{extendedgap}Inclusion of neighboring domains: a complete basis}

There is a need to extend the single domain basis because it does not
properly capture the long time behavior of $C(t)$ for $c$ less than
0.5, as seen from Fig.~\ref{Mathematicaresults}. The lack of
quantitative agreement between theory and simulation at long times
implies essentially that there is important slow behavior in the
memory function $\mathsf M^{D}(t)$ given in Eq.~(\ref{MD11}) that cannot
be neglected.

This situation is reminiscent of that in fluids. There, one starts out
describing time correlation functions in terms of the \emph{linear}
dependence on the hydrodynamic fields of mass, momentum and
energy density, i.e., one takes these to comprise the set $A$ of
Sec.~\ref{modecoupling}\cite{Keyes77,Zwanzigbook}.  But at lower
temperatures or higher densities this does not suffice because $\mathsf
M^D(t)$ turns out to no longer be a fast decaying function.  To fix
this situation, \emph{i.e.}, to represent the missed slow behavior in
$\mathsf M^D(t)$, one needs to augment the linear basis by vectors
orthogonal to it. For this, one can take products of $A$ (with proper
subtractions to assure orthogonality); these additional basis vectors
are called \emph{multi-linear
modes}\cite{MachtaOppenheim82,KavassalisOppenheim88}. The coupling of
the linear modes to the multi-linear ones ``renormalizes'' the bare
values of $\mathsf M^D$ found using only $A$ to $\mathsf M^D+\tilde{\mathsf
\Sigma}(z)$, where $\tilde{\mathsf\Sigma}$ is a \emph{self-energy}. The
$z$ dependence of this self-energy is such that it can describe slowly
decaying behavior such as long-time tails. Since the multi-linear
modes can be interpreted as products of linear hydrodynamics modes,
this procedure amounts to a nonlinear coupling of hydrodynamic modes
and is hence called mode-coupling theory.

Similarly, if for the East model, the matrix $\mathsf M^{E}$ is taken to
be represented at the linear level by Eq.~\eqref{M11Egap}, where the
linear basis (using analogous nomenclature as above) composing the set
of slow variables is taken to be $A=\{\hat Q_0, \hat Q_1\}$, then the
memory function corresponds to an infinite square matrix represented
at the linear-linear level that effectively renormalizes the matrix
elements of $\mathsf M^{E}$ to $\mathsf M^E+\tilde{\mathsf M}^D$, according
to Eq.~\eqref{laplacesol}. Hence, $\tilde{\mathsf M}^D$ takes the role
of the self-energy here, and must describe contributions to the decay
of the spin-spin correlation function due tos the projection of the
dynamical evolution onto a space orthogonal to the linear basis set
$A$.  In other words, the single domain basis does not span the
ergodic component and fails to capture all the slow dynamics of the
spin fluctuation variable $\hat{Q}_{0}$.  To represent the missed slow
evolution, the single domain basis set must be expanded to include
additional slow modes and their coupling to the linear modes must be
computed.

To deduce the appropriate extension of the basis, it is helpful to
realize, as Fig.~\ref{Mathematicaresults} shows, that the decay of
$C(t)$ that is predicted by the extended linear basis, $\{\hat Q_0,
\hat Q_1\}$, is too rapid. A reasonable explanation of this particular
deviation is that in the single domain basis, the final spin $n_{k+1}$
in Eq.~\eqref{Akdef} decays regardless of the spin configuration to
its right. As a result, any slowing down effect of a persistent
down-spin domain to the right of $n_{k+1}$ is missed.  It therefore
seems natural to try to fix the too rapid decay of the $C(t)$ by
augmenting the basis with a second down-spin domain,
\begin{eqnarray}
\hat Q_2(l_1,l_2) &=& \frac{1}{Z_2(l_1,l_2)}(n_0 - c)
	\prod_{j_1=1}^{l_1}(1-n_{j_1})
		n_{l_1+1}
\nonumber
\\&&\times
	\prod_{j_2=l_1+2}^{l_1+l_2+1}(1-n_{j_2})
		(n_{l_1+l_2+2}-c),
\label{defBk}
\end{eqnarray}	
which carries an index doublet $(l_1,l_2)$ of which each member
$l_j$ can take integer values from zero to infinity, and
\begin{equation}
  Z_2(l_1,l_2) = c^{3/2} (1-c)^{1+[l_1+l_2]/2}.
\label{theno}
\end{equation}
As before, unnormalized versions
$Q_2(l_1,l_2)=Z_2(l_1,l_2)\hat
Q_2(l_1,l_2)$ are defined as well:
\begin{eqnarray}
Q_2(l_1,l_2) &=& (n_0 - c)
	\prod_{j_1=1}^{l_1}(1-n_{j_1})
		n_{l_1+1}
\nonumber
\\&&\times
	\prod_{j_2=l_1+2}^{l_1+l_2+1}(1-n_{j_2})
		(n_{l_1+l_2+2}-c),
\label{unno}
\end{eqnarray}
Diagrammatic representations of the $Q_2(l_1,l_2)$ are shown in
Table~\ref{Bdiagrams}. These are orthogonal to $Q_0$, as well as to
$Q_1$, since in a diagrammatic representation of
$\inner{Q_0}{Q_2(l_1,l_2)}$, the trailing open circle
($n_{l_1+l_2+2}-c$) of $Q_2$ is not covered, yielding zero, and in the
diagram of $\inner{Q_1(k)}{Q_2(l_1,l_2)}$, it is impossible to line up
the trailing open circles of $Q_1(k)$ and $Q_2(l_1,l_2)$ without
having the solid dot $\bullet$ (=$n$) overlap with a horizontal line
$\textdline{2}{\cbar}$ (=$1-n$), which yields zero
[cf.~Table~\ref{rules}]. It is easy to establish that the $\hat
Q_2(l_1,l_2)$ are also orthonormal among themselves.

\begin{table}[tb]
\hrule
\begin{tabular}{c@{\ \ \ \ }c}
\dline{4}{\setO\setX\setO}
& 
\dline{8}{\setO\cbar\txt{l_1}\cbar\cbar\setX\setO}
\\ 
$Q_2(0,0)=R(0)$ 
&
$Q_2(l_1,0)=R(l_1)$
\\
\dline{6}{\setO\setX\cbar\txt{l_2}\cbar\setO}
&
\dline{10}{\setO\cbar\txt{l_1}\cbar\cbar\setX\cbar\txt{l_2}\cbar\setO}
\\ 
$Q_2(0,l_2)=S(0,l_2-1)$ & 
$Q_2(l_1,l_2)=S(l_1,l_2-1)$ 
\end{tabular}
\hrule
\caption{Diagrams of the extension of the domain basis. (The notation
using $R$ and $S$ is used only in the appendix.)}\label{Bdiagrams}
\end{table}        

There is no obvious reason to stop this procedure at the two-domain,
or ``bi-linear'', level and, in fact, the basis can be extended to a
complete set in the relevant ergodic component in a straightforward
fashion. The elements of this complete basis are most easily written
diagrammatically as a sequence of $\alpha$ down-spin domains of
different sizes $k_j$, separated by single up-spins:
\begin{equation}
  \hat Q_\alpha(k_1,\ldots,k_\alpha)
  = \frac{1}{Z_\alpha(k_1,\ldots,k_\alpha)}
\dline{12}{\setO\cbar\txt{k_1}\cbar\cbar\setX\cbar\txt{k_2}\cbar\cbar\setX}
\cdots
\dline{7}{\setX\cbar\txt{k_\alpha}\cbar\cbar\setO}.
\label{Qdef}
\end{equation}
Here, $\alpha=0\ldots\infty$, $k_j=0\ldots\infty$ $(j=1\ldots\alpha)$
and
\begin{equation}
  Z_\alpha(k_1,\ldots,k_\alpha)
  = \left\{\!\begin{array}{ll}
c^{1/2}(1-c)^{1/2} & \text{ if } \alpha=0\\
c^{(1+\alpha)/2}(1-c)^{1+\sum_{j=1}^\alpha k_j/2}&\text{otherwise.}
\end{array}
\right.
\end{equation}
It is easy to see that the $\hat Q_\alpha$ are all independent: the
inner product of two of them is zero unless their diagrams [as in
Eq.~\eqref{Qdef}] are equally long, so that both open circles are
covered. But then the interior of the diagrams has to have matching
top and bottom parts too, otherwise a solid dot and a horizontal line
occur at the same site, and this gives zero (Table~\ref{rules}). The
only nonzero inner product of a $\hat Q_\alpha(k_1,\ldots k_\alpha)$
is therefore with itself.  Due to our choice of normalization
$\average{[\hat Q_\alpha(k_1,\ldots,k_\alpha)]^2}=1$.  Since each
$\hat Q_\alpha$ is orthonormal to all others, each contributes a
unique direction in the Hilbert space to the basis which could not be
formed from the others: The $\hat Q_\alpha$ are independent.

To also establish completeness, we will now count the number of
element of the above basis in the finite system of $N$ spins. In that
system, there exist only $\hat Q_\alpha$ for which
$\alpha+1+\sum_{j=1}^\alpha k_j\leq N$ (which also limits
$\alpha<N$). Elementary combinatorics shows that the number of
different $\hat Q_\alpha$ for given $\alpha$ and $N$ is
$\binom{N-1}{\alpha}$. The total number of $\hat Q_\alpha$ is thus
$\sum_{\alpha}^{N-1}\binom{N-1}{\alpha}=2^{N-1}$.  Thus the above set
of $2^{N-1}$ basis vectors covers only half of the full Hilbert space,
which has $2^N$ dimensions.  But it is easy to see which basis vectors
are missing and why they are not important. The expression in
Eq.~\eqref{Qdef} always starts with $n_0-c$, even though the first
spin can have two values. Independent vectors can be found by taking a
different expression for the first spin. In fact, one can take $1$,
i.e., one could consider variants of the basis vectors in
Eq.~\eqref{Qdef} in which the factor of $(n_0-c)$ is not present. Call
these $\check Q_\beta$, of which there are as many as there are $\hat
Q_\alpha$. These new vectors are orthogonal to each other as well as
to the $\hat Q_\alpha$ because in $\average{\hat Q_\alpha\check
Q_\beta}$ the initial $\circ$ in the diagram of $\hat Q_\alpha$ is not
covered by the diagram of $\check Q_\beta$ and $\average{\circ}=0$.
Thus, the $\check Q_\beta$ are the missing basis vectors.  However,
they are completely unimportant here because the $\check Q_\beta$ are
also orthogonal to $\mathcal L\hat Q_\alpha$, as is seen from the fact
that $\mathcal L\hat Q_\alpha$ will always have a ``$\circ$'' as a
first element ($\mathcal L
\textdline{2}{\setO}=-\textdline{3}{\setO\setX}$) so that the
inner product with $\check Q_\beta$ is zero (for this it is important
\emph{not} to have periodic boundary conditions).

So the basis set $\hat Q_\alpha$ is not a complete basis for all
possible spin configurations, but it \emph{is} a complete orthonormal
basis for all spin configurations to which the $\hat Q_\alpha$ couple.

These considerations also imply that the East model is not ergodic:
the state space contains at least two ergodic components, which are
such that a configuration in one of them can never make a transition
to any configuration in the other. Noting that the space spanned by
$\check Q_\beta$ contains all quantities insensitive to the value of
$n_0$, one realizes that it constitutes an East model with an
effective length of $N-1$. The argument above then shows that the
state space of this smaller East model can also be split into at least
two ergodic components. Applying this argument recursively reveals
that there are $N+1$ ergodic components. The $p$-th ergodic component
consists of functions not sensitive to the values of spins $n_0$
through some $n_{p-1}$, with $0\leq p\leq N$, and has $2^{N-p-1}$
dimensions if $p<N$ and one dimension if $p=N$. The collection of
these ergodic components has $1+\sum_{p=0}^{N-1} 2^{N-p-1}=2^N$
dimensions, and thus indeed spans the full Hilbert space of the spin
chain of length $N$.

Since we are interested in the time auto-correlation function of spin
$n_0$, the relevant ergodic component is the one spanned by $\hat
Q_\alpha$, and we conclude that the $\hat Q_\alpha$ are the only basis
vectors needed. Having established the ``relevant completeness'', one
can take the limit $N\to\infty$ again, so we need not worry about the
boundary condition imposed on $n_{N}$.

The extension of the basis set to include an arbitrary number of
domains is useful in developing a systematic approach to generate
successive improvements for $\tilde C$ for lower $c$.  Since the basis
$A=\{\hat Q_\alpha\}$ spans the ergodic component of $\hat{Q}_0$, it
follows that $\varphi(t) = e^{(1-\mathcal P)\mathcal L t}(1-\mathcal
P)\mathcal L \hat{Q}_0 = 0$ and the memory function $\mathsf M^{D}(t)$
vanishes.  Hence, from Eq.~\eqref{laplacesol}, $\tilde{\mathsf
G}(z) = \left( z\identity - \mathsf M \right)^{-1}$, where the full
matrix $\mathsf M=\mathsf M^E$ in this complete basis can be written as
\begin{equation}
  \mathsf M = \left[\begin{matrix}
              \mathsf M_{00} & \mathsf M_{01} \\
              \mathsf M_{01}^\dagger & \mathsf M_{11} & \mathsf M_{12} \\
                           & \mathsf M_{12}^\dagger & \mathsf M_{22} & \ddots\\
                           &              &    \ddots       & \ddots
             \end{matrix} \right]
\end{equation}
where $\mathsf M_{\alpha\beta}=\mel{\hat Q_\alpha}{\mathcal L}{\hat
Q_\beta}$ and it was used that this is zero unless $|\alpha-\beta|<2$,
as can easily be shown (also, diagonal dots do not denote repetition
now: all $\mathsf M_{\alpha\beta}$ can be different).  Generalizing
Eq.~\eqref{contfrac} by repeatedly applying the matrix equality (see
e.g.\ Ref.~\cite{NumericalRecipes}, page 70)
\newcommand{\ct}{\mathsf d}
\begin{equation}
\begin{bmatrix}
\mathsf a&\mathsf c\\
\ct&\mathsf b
\end{bmatrix}^{-1}=
\begin{bmatrix}
[\mathsf a- \frac{\mathsf c\,\mathsf\ct}{\mathsf b}]^{-1}
&
-[\mathsf a - \frac{\mathsf{c}\,\mathsf\ct}{\mathsf b}]^{-1}
\mathsf c \mathsf b^{-1}
\\&\\
-[\mathsf b-\frac{\ct\,\mathsf c}{\mathsf a}]^{-1}
\ct\mathsf a^{-1}
&
[\mathsf b-\frac{\ct\,\mathsf c}{\mathsf a}]^{-1}
\end{bmatrix},
\label{splitinverse}
\end{equation}
one finds
\begin{eqnarray}
  \tilde{\mathsf C}(z) &=& \left[\begin{matrix}
              z\identity-\mathsf M_{00} & -\mathsf M_{01} \\
              -\mathsf M_{01} & z\identity-\mathsf M_{11} & -\mathsf M_{12}\\
                           & -\mathsf M_{12}^\dagger & z\identity-\mathsf M_{22} &
\ddots \\
                           &              &          \ddots       & \ddots
             \end{matrix} \right]^{-1}_{11}
\label{formal0}\\
&=&\cfrac{1}{z\identity-\mathsf M_{00}-\cfrac{\mathsf M_{01}\,\mathsf
      M_{01}^\dagger}{z\identity-\mathsf M_{11}-\cfrac{\mathsf M_{12}\,\mathsf
        M_{12}^\dagger}{z\identity-\mathsf M_{22}-\ldots}}} 
\label{formal}\\
&=& \cfrac{1}{z\identity-\mathsf M_{00}-\cfrac{\mathsf M_{01}\,\mathsf
      M_{01}^\dagger}{z\identity-\mathsf M_{11}-\tilde{\Sigma}_{11}(z)}} ,
\end{eqnarray}
where the self-energy matrix at the linear-linear level is defined to
be
\begin{eqnarray}
\tilde{\Sigma}_{11}(z) &=& \cfrac{\mathsf M_{12}\,\mathsf
        M_{12}^\dagger}{z\identity-\mathsf M_{22}-
	\cfrac{\mathsf M_{23}\,\mathsf M_{32}^\dagger}{z\identity -
        \mathsf M_{33} - \ldots}} 
\\
&=& \cfrac{\mathsf M_{12}\,\mathsf
        M_{12}^\dagger}{z\identity-\mathsf M_{22}-\tilde{\Sigma}_{22}(z)}.
\label{defSigma11}
\end{eqnarray}
For convenience a non-standard (but unique) notation for a matrix
fraction has been introduced here, such that if $\mathsf A$, $\mathsf
B$ and $\mathsf C$ are matrices then
\begin{equation}
\cfrac{\mathsf A\,\mathsf B}{\mathsf C} 
\equiv
 \mathsf A\cdot\mathsf C^{-1}\cdot\mathsf B.
\end{equation}  
This notation saves a lot of space and avoids many nested parentheses
and inverses that would be required in more standard notation.  We
remark that Eq.~\eqref{formal0} is similar to the matrix formalism of
Wu and Cao\cite{WuCao04}, while Eq.~\eqref{formal} has similarities to
the continued fraction formalisms of Mori\cite{Mori65a,Mori65b} and
Schneider\cite{Schneider76}.  The structure of Eq.~\eqref{formal0} is
that of a mode-coupling theory in which the role of mode order is
played by the number of domains.  The effect of the higher order modes
is to renormalize the ``transport'' coefficients approximated by
$\mathsf M^{E}$ at the linear level.

By truncating Eq.~\eqref{formal} at ever deeper levels, i.e.,
setting $\mathsf M_{\alpha,\alpha+1}=0$, corresponding to
$\tilde{\Sigma}_{\alpha \alpha}(z) = 0$, for increasing $\alpha$ one
gets expressions which work well for ever lower values of $c$, denoted
as $\tilde C^{(\alpha)}(z)$. For example, truncating at the zeroth
level gives
\begin{equation}
\tilde  C^{(0)}(z) = \frac{1}{z-\mathsf M_{00}}= \frac{1}{z+c},
\label{zdr}
\end{equation}
while truncating at the first, linear, level yields the result in
Eq.~\eqref{Cz}.  Following this procedure further, the first
correction to the linear basis results involves evaluating the
self-energy in the approximation where one ignores the effects of
three-domains and higher, i.e. [cf. Eq~\eqref{defSigma11}],
\begin{equation}
\tilde{\Sigma}_{11}(z) \approx \cfrac{\mathsf M_{12}\,\mathsf
        M_{12}^\dagger}{z\identity-\mathsf M_{22}},
\label{Sigma11approx}
\end{equation}
corresponding to a bi-linear type of mode-coupling theory. Due to the
simplicity of the coupling with respect to mode order and for
different domain sizes, an exact expression for this approximate
self-energy can be obtained.  In the appendix 
$\tilde{\Sigma}_{11}(z)$ is explicitly evaluated to be:
\begin{equation}
  \tilde{\mathsf\Sigma}_{11} =
 \left[\begin{matrix}
  \tilde\eta_1
 &-\sqrt{1-c}\,\tilde\eta_1\\
  -\sqrt{1-c}\,\tilde\eta_1
 & (1-c)\tilde\eta_1
   +\tilde\eta_2
  &-\sqrt{1-c}\,\tilde\eta_2\\
  &-\sqrt{1-c}\,\tilde\eta_2
  & (2-c)\tilde\eta_2
  &\ddots\\
  &&\ddots&\ddots
  \end{matrix}\right]
\label{Sigma}
\end{equation}
where the functions $\tilde\eta_j$ are given by Eqs.~\eqref{aj} and
\eqref{ej}.  Using this expression for the self-energy, the
linear-linear matrix $\tilde{\mathsf G}(z)$ of the previous section
(which is in fact the top-left block of the inverse matrix on the
right hand side of Eq.~\eqref{formal0} incorporating the zeroth and
first level) is renormalized to
\begin{widetext}
\begin{eqnarray}
\tilde{\mathsf G}^{(2)}_R &=&
\left[
\begin{matrix}
  z \identity - \mathsf M^E - \begin{pmatrix}
0&0\\
0&\tilde{\mathsf\Sigma}_{11}
\end{pmatrix}
\end{matrix}
\right]^{-1}
\label{renormG} 
\\ 
&=&
\left[\begin{matrix}
z+c     
&\sqrt{c(1-c)}
\\
\sqrt{c(1-c)}
&z+1-\tilde\eta_1
&-\sqrt{1-c}(c-\tilde\eta_1)
\\
&-\sqrt{1-c}(c-\tilde\eta_1)
&z+(2-c)(c-\tilde\eta_2)+(1-c)(\tilde\eta_2-\tilde\eta_1)
&-\sqrt{1-c}(c-\tilde\eta_2)
\\
&&-\sqrt{1-c}(c-\tilde\eta_2)
& z+(2-c)(c-\tilde\eta_2)
&\ddots
\\
&&&\ddots&\ddots
\end{matrix}\right]
^{-1} 
\end{eqnarray}
from which the single spin time correlation function
$\tilde{C}^{(2)}=[\tilde{\mathsf G}^{(2)}_R]_{11}$ is computed with
the continued fraction expression in Eq.~\eqref{contfrac} to be
\begin{equation}
\tilde{C}^{(2)} = \cfrac{1}{z+c-\cfrac{c(1-c)}{\alpha (c,z)}}
\label{Cprelex}
\end{equation}
where $\alpha(c,z)$ is 
\begin{equation}
\alpha (c,z) = z+1
-\tilde\eta_1
-\cfrac{(1-c)(c-\tilde\eta_1)^2}
{ z+(2-c)(c-\tilde\eta_2)+  (1-c)(\tilde\eta_2- \tilde\eta_1) - \tilde\gamma^{(2)}}.
\label{defAlpha}
\end{equation}
In Eq.~\eqref{defAlpha}, the repetitive part $\tilde\gamma^{(2)}$ satisfies
\begin{equation}
  \tilde\gamma^{(2)}
  =
  \cfrac{(1-c)(c-\tilde\eta_2)^2}
   {z+(2-c)(c-\tilde\eta_2)
  -\tilde\gamma^{(2)}}
.
\label{tildegammaeq}
\end{equation}
This is solved by
\begin{equation}
\tilde\gamma^{(2)} = 
\frac12\left\{z
+
(2-c)(c-\tilde\eta_2)
-\sqrt{4(c-\tilde\eta_2)z
+(z-c(c-\tilde\eta_2))^2
}
\right\}.
\label{gammatwo}
\end{equation}
Note the resemblance with $\tilde\gamma^{(1)}$ in Eq.~\eqref{gammasol}. 
\end{widetext}

\begin{figure}[t]
\centerline{\includegraphics[angle=-90,width=0.47\textwidth]{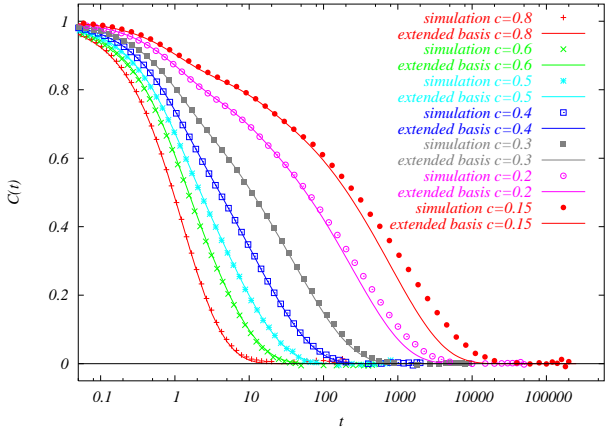}}
\caption{Results for the single spin time correlation function $C(t)$
using the extended basis $\{\hat Q_0,\hat Q_1,\hat Q_2\}$ in
Eq.~\eqref{Cprelex} of Sec.~\ref{extendedgap} (numerically Laplace
inverted using {\it Mathematica}), compared to simulation data.}
\label{Mathematicaxresults}
\end{figure}

As before, the exact result in Eq.~\eqref{Cprelex} is inverted
numerically using Stehfest's
algorithm\cite{Stehfest70a,Stehfest70b,StehfestMathematicaPackage}.
For various values of $c$, the results are shown in
Fig.~\ref{Mathematicaxresults} and compared with data from simulations
of the East model. Notice that there is a huge improvement over the
results obtained using only the single domain basis $\{\hat Q_0,\hat
Q_1\}$ in Fig.~\ref{Mathematicaresults}.  There is now excellent
agreement between the theory and the data for $c\geq 0.4$, and
reasonable agreement up to $c\approx 0.3$. Furthermore, while the
theoretical decay is still too fast for $c<0.3$, the small time
behavior is captured beautifully. In particular, the shoulder that
appears in the simulations for low $c$ is reproduced by the extended
theory as well, something the single domain basis could not do.

In the low temperature region (small $c$), the long time behavior of
the spin autocorrelation function predicted by the two-domain basis set
is well-described by a stretched exponential $C(t) \sim \exp{[-\left(
t/\tau \right)^{\beta}]}$ with a temperature independent
stretching exponent of $\beta \approx 0.6$.  Although it is
encouraging that the stretched exponential time profile is indeed
predicted by the theory, simulations indicate that in fact the
stretching exponent $\beta$ should have a weak temperature
dependence\cite{BerthierGarrahan03}, with $\beta$ decreasing in value
as the temperature decreases.  The origin of this discrepancy between
our theory and numerical simulation is not clear and is under
investigation.

In principle, the effect of three down-spin domains (tri-linear modes)
can be included in the same spirit, i.e., by evaluating the
self-energy at the two-domain level $\tilde{\Sigma}_{22}(z)$ using
matrix methods similar to those applied to obtain Eq.~\eqref{Sigma}.
Unfortunately, the algebra becomes even more cumbersome and explicit
evaluation of the self-energy matrices at higher and higher order
becomes effectively impossible.  Alternatively, one can resort to
numerical approaches in which the maximum domain size $k_m$ is fixed
and all matrix inversions are carried out numerically.  By monitoring
convergence to a set level of precision, such a procedure provides a
systematic and numerically tractable method of predicting the decay of
the spin autocorrelation function for arbitrary values of $c$.

\section{\label{relax}Relaxation behavior}

One of the main advantages of the matrix method outlined here is that
it is straightforward to obtain analytic predictions for rather
detailed features of the dynamics.  For example, one of the commonly
calculated quantities from simulation data is the relaxation time
$\tau$.  For systems exhibiting such non-trivial relaxation behavior
as stretched-exponential, the definition of the relaxation time is a
matter of choice.  Perhaps the most sensible way to view the
relaxation time for such systems is to consider it as the
weighted-average of a distribution of relaxation times.  For example,
based on the spectral decomposition of the Liouville operator, one can
formally write the spin autocorrelation function as a weighted sum of
exponentials with relaxation times $\tau_n$,
\begin{equation}
C(t)= \sum_n  c_n \exp(-t/\tau_n).
\label{Csum0}
\end{equation}
Since the Liouville operator is Hermitian and the spin variables are
real, one is guaranteed that the relaxation times $\tau_n$ and
coefficients $c_n=\inner{\hat{Q}_0}{\psi_n}^2$, where $\big| \psi_n
\rangle$ are the right (and left) eigenvectors of the Liouvillian
${\cal L}$, are real and positive.  Furthermore, since
$C(t=0)=1=\sum_n c_n$, the coefficients $c_n$ are proper weights for
the relaxation time $\tau_n$.  However, since ${\cal L}$ is of
infinite dimension, its spectrum can be (partially or completely)
continuous, so the more general expression to replace
Eq.~\eqref{Csum0} is
\begin{equation}
  C(t)= \int  \rho(\tau') \exp(-t/\tau')\,d\tau',
\label{Csum}
\end{equation}
where  $\rho(\tau')\geq0$, $\rho(\tau'<0)=0$ and $\int
\rho(\tau')\,d\tau'=1$.  One can therefore define the {\it average}
relaxation time as
\begin{equation}
\tau = \int \rho(\tau')\tau' \,d\tau'.
\label{averageRelaxationTime}
\end{equation}
Noting that the Laplace transform $\tilde{C}(z)$ of Eq.~\eqref{Csum} is
\begin{equation}
\tilde{C}(z) = 
\int \frac{\rho(\tau')}{z+1/\tau'}d\tau',
\end{equation} 
we see that 
\begin{equation}
  \tau = 
 \int \rho(\tau')\tau' \,d\tau'
 = \tilde{C}(z=0).  
\label{relaxfromCz}
\end{equation}
Note that in the case in which a single relaxation time $\tau^*$
dominates all others, one observes that $\tau \approx \tau^*$ since
$\rho(\tau')\approx \delta(\tau'-\tau^*)$.

Note also that in taking the point $z=0$, the expression is sensitive
to long time behavior. This in contrast to e.g. the average rate $\int
\rho(\tau') (1/\tau') d\tau'$ which by Eq.~\eqref{Csum} is just
$-(d/dt)C(t=0)=c$ and contains no information on the long time
behavior.

\begin{figure}[t]
\centerline{\includegraphics[width=0.43\textwidth]{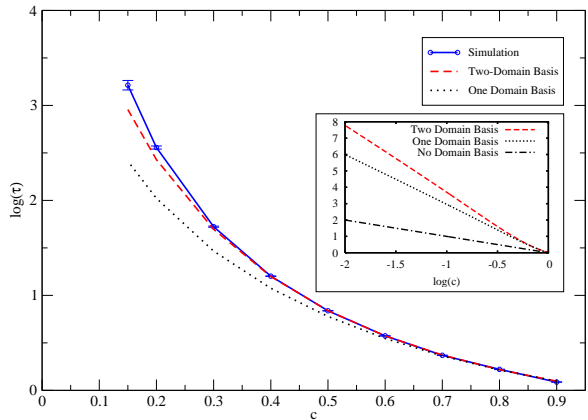}}
\caption{Logarithm (base 10) of average relaxation time $\tau$ for
various values of $c$. The inset shows the same as a function of the
logarithm of $c$ (also base 10) for the theoretical results, with
slopes of $-1$, $-3$ and $-4$ showing their scaling behavior. }
\label{averageRelax}
\end{figure}

Given the analytical results for the Laplace transform of the spin
autocorrelation function in the one-domain [Eq.~\eqref{Cz}] and
two-domain representations [Eq.~\eqref{Cprelex}] of the slow dynamics,
explicit expressions for $\tau (c)$ can be obtained by setting $z=0$
in the respective equations.  For example, not including domains as in
Eq.~\eqref{zdr} gives $\tau^{(0)}=1/c$, while in the one-domain basis,
Eqs.~\eqref{Cz} and \eqref{relaxfromCz} lead to the simple result
\begin{equation}
\tau^{(1)}  = \frac{1-c+c^2}{c^3},
\label{relaxTimeSimple}
\end{equation}
in which the average relaxation time diverges as $c^{-3}$ as the
concentration $c$ approaches zero. Furthermore, in the two-domain
representation, the average relaxation time is a complicated function
of $c$.  In the limit that $c \rightarrow 0$, we find that $\tau^{(2)}
\sim c^{-4}$.  In Fig.~\eqref{averageRelax}, the theoretical
predictions of the average relaxation time in the one-domain and
two-domain basis sets are compared with numerically-integrated simulation
data.  Note that as is evident from Figs.~\eqref{Mathematicaresults}
and \eqref{Mathematicaxresults}, the two-domain predictions
significantly improve the one-domain results but still underestimate
the relaxation time of the system at small values of $c$.

From the relationship between $c$ and $\beta \mu$ in Eq.~\eqref{cbm},
it is clear that at low temperatures $c \sim \exp(-\beta \mu)$. Since
the logarithm of the average relaxation time $\log(\tau)$ is
proportional to $\log(1/c)$ for $\tau^{(0)}$, $\tau^{(1)}$ and
$\tau^{(2)}$, a plot of $\log(\tau)$ versus $\beta \mu$ yields a
straight line in the small $c$ (low temperature) limit.  Thus we can
conclude that the zero, one and two-domain basis sets all
yield a relaxation time that diverges according to the Vogel-Fulcher
law $\tau \sim \exp\{ -\text{const}/(T-T_0)\}$ with a glass transition
temperature of $T_0=0$.  Note that these results are in contrast with
the exact result for the equilibration time $\tau_e$ of a system
quenched to $T=0$ where $\tau_e \sim \exp\{ \text{const}/T^2
\}$\cite{SollichEvans99,AldousDiaconis02}.  This finding is somewhat
surprising given that the equilibration time was calculated in the
asymptotic small $c$ regime using ideas of domain structure rather
similar to those presented here.

\begin{figure}[t]
\centerline{\includegraphics[width=0.5\textwidth]{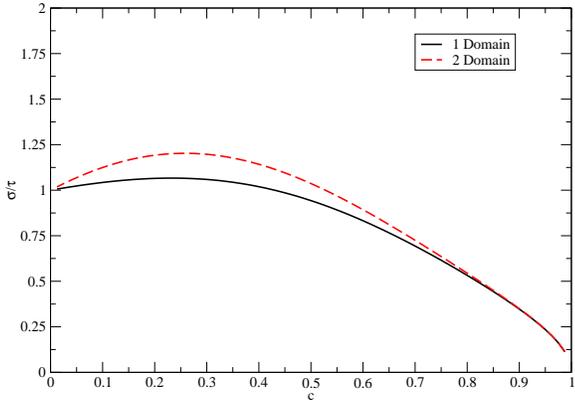}}
\caption{Ratio $\sigma/\tau$ of the width $\sigma$ of relaxation time
distribution to the average relaxation time $\tau$ as a function of
$c$, in the one-domain and two-domain bases.}
\label{widthRelax}
\end{figure}

Given the relatively simple structure of the matrix $\tilde{G}(z)$, it
is easy to numerically examine many detailed features of the
relaxation given a finite domain basis set specified by setting a
maximum domain size $k_m$.  For example, one can easily examine how
the spectrum of ${\cal L}$ depends on $c$.  At the same time, the
actual distribution of the $c_n$ can be computed numerically to see
how many relaxation modes are relevant as a function of $c$.  From
this information, one can try to attempt to establish a link between
the distribution of relaxation times as a function of temperature and
the asymptotic stretched exponential form, as suggested in
reference~\cite{Palmeretal84}.  This may be an instructive way to
examine the failure of the two-domain basis to correctly predict the
temperature dependence of the stretching exponent $\beta$.  However
since analytical results are available for all quantities, it is
desirable to obtain analytical expressions for such features as the
\emph{width} or spread $\sigma$ of the relaxation times $\tau'$ as a
function of $c$. The spread in $\tau'$ is defined by
\begin{equation}
    \sigma = \sqrt{\int \rho(\tau')(\tau'-\tau)^2\,d\tau'}. 
\end{equation}
Now one can use that $\tilde{C}'(0) = \lim_{z \rightarrow 0} d/dz
\,\tilde{C} (z) = - \int \rho(\tau') {\tau'}^2d\tau'$ [cf.\
Eq.~\eqref{relaxfromCz}] to write $\sigma = {\{ - \tilde{C}'(0) -
[\tilde{C}(0)]^2 \}^{1/2}}$.  Since we have obtained closed expressions for
$\tilde{C}(z)$, analytic expressions can be obtained for $\sigma$.
For example, using the one-domain basis set, we find that
\begin{equation}
\sigma^{(1)} = \frac{\sqrt{1-c}}{c^3},
\label{width}
\end{equation}
whereas the expression for $\sigma^{(2)}$ in the two-domain basis is a
complicated function of $c$ [note: $\sigma^{(0)}$ is actually zero].
From these analytical expressions for $\sigma$, one immediately sees
that, in fact, $\sigma$ diverges as $c$ approaches zero in the
\emph{same} way as $\tau$.  Hence a plot of $\sigma/\tau$ remains
finite for all values of $c$.  Furthermore, noting that in the
one-domain basis,
\begin{equation}
\sigma^{(1)} / \tau^{(1)} = \frac{\sqrt{1-c}}{1-c+c^2},
\end{equation}
it is evident that $\lim_{c\rightarrow 0} \sigma^{(1)} / \tau^{(1)} =
1$.  Surprisingly, the same conclusion holds in the two-domain basis,
as is evident from Fig.~\ref{widthRelax}.  Note that at large
values of $c \approx 1$, $\sigma \approx 0$ indicating that the
relaxation is dominated by a single mode.

We note that higher order derivatives of $\tilde C(z)$ at $z=0$ can
similarly be used to investigate further characteristics of the
relaxation time distribution such as the skewness and the
kurtosis. More generally, Eq.~\eqref{Csum} shows $C(t)$ to be the
Laplace transform of 
the distribution of relaxation rates. If $r=1/\tau$ are the relaxation
rates, then their distribution is $P(r)=r^{-2}\rho(1/r)$ and
Eq.~\eqref{Csum} can be written as
\begin{equation}
    C(t) = \int_0^\infty  P(r) \exp(-rt)\,dr.
\end{equation}
In this sense, $C(t)$ is the Laplace transform of $P(r)$.  Thus, given
$C(t)$, one might expect to be able to use the numerical Laplace
inverse of the Stehfest algorithm to obtain $P(r)$. Unfortunately it
turns out that using Stehfest's numerical Laplace inverse method on
$C(t)$, which was itself obtained from $\tilde C(z)$ by the same
method, is unstable; it in fact yields an incorrect result for $P(r)$,
namely a highly oscillating function, which is not non-negative and
not normalized to one.  Since the distribution of relaxation rates was
considered here mainly as an illustration of the power of the
theoretical approach presented in this paper, solving the numerical
instability associated with applying the Stehfest algorithm twice and
determining the distribution of relaxation rates in detail, is left
for future work.

\section{\label{higherorder}Higher order correlation functions}

\begin{figure}[t]
\centerline{\includegraphics[width=0.43\textwidth]{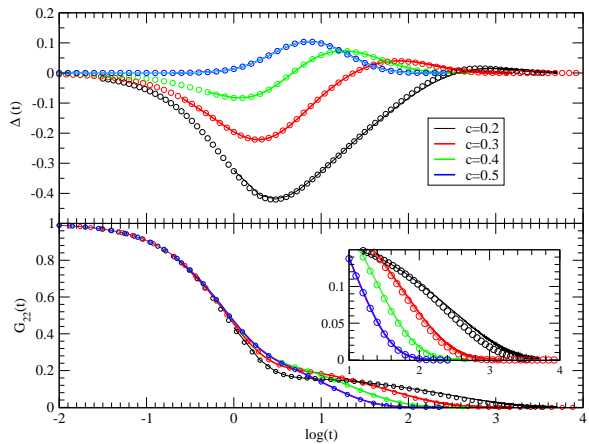}}
\caption{Higher order correlation functions. Top: The non-Gaussian
measure $\Delta (t)$ defined in Eq.~\eqref{defDelta}. Bottom: The
neighbor-pair auto-correlation function $G_{22}(t)$.  In both graphs,
open circles and solid lines correspond to \emph{theoretical} and
\emph{simulation} values, respectively.}
\label{fourPoint}
\end{figure}

Given the glassy nature of the dynamics of the East model, it is
interesting to probe higher order correlation functions to examine
issues of cooperativity in the dynamics and non-Gaussian
statistics. In particular, one can look at the neighbor-pair spin
correlation function
\begin{eqnarray}
\average{\hat{n_i}(t)\hat{n}_{i+1}(t) \hat{n}_i(0) \hat{n}_{i+1}(0)}
&=& \average{\hat{Q}_1(k;t)\hat{Q}_1(k;0)} \delta_{k,0} 
\nonumber
\\
&=& G_{22}(t),
\label{defG22}
\end{eqnarray}
and a related quantity
\begin{equation}
\Delta (t) = G_{22}(t) - \average{\hat{n}_{i}(t)\hat{n}_i(0)} \average{\hat{n}_{i+1}(t)\hat{n}_{i+1}(0)}
\label{defDelta}
\end{equation}
that examines the non-Gaussian nature of the normalized spin
fluctuation variable $\hat{n}_i$.  Given the simplicity of the matrix
method, it is relatively straightforward to obtain analytic
expressions for higher order correlation functions such as
Eq.~\eqref{defG22}.  For example, from the definition of the
neighbor-pair spin variable, which corresponds to the linear basis set
element $\hat{Q}_{1}(0)$, it follows that the Laplace transform
$\tilde{G}_{22}$ of the function $G_{22}(t)$ is the $2-2$ element of
the infinite matrix $\tilde{\mathsf G}$, which, in the two-domain basis
approximation, is given by Eq.~\eqref{renormG}.  Using standard matrix
inversion methods, the $2-2$ element of $\tilde{\mathsf G}^{(2)}$ is
\begin{equation}
\tilde{G}_{22}^{(2)} = \cfrac{1}{\alpha (c,z) - \frac{c(1-c)}{z+c}} ,
\label{G22}
\end{equation}
where $\alpha (c,z)$ is given in Eq.~\eqref{defAlpha}.  In
Fig.~\eqref{fourPoint}, the functions $G_{22}(t)$ and $\Delta(t)$ are
plotted versus time for various values of $c$ (using Stehfest's
algorithm for the inverse Laplace transform).  Note that the agreement
between the theoretical predictions and the simulation data is
excellent for all times for all but the smallest value $c=0.2$.

The neighbor-pair autocorrelation function exhibits several remarkable
properties that are rather unlike those of the spin autocorrelation
function. Note that in the short time limit $t \leq 1$ the relaxation
of $G_{22}(t)$ is \emph{independent} of the equilibrium up-spin
concentration $c$.  This result can be explained by examining a short
time expansion (large $z$) of $\tilde{G}_{22}$, from which it is seen
that $\alpha (c,z) \sim z+1$ and hence $\tilde{G}_{22} \sim 1/(z+1)$,
corresponding to simple exponential relaxation $G_{22}(t) \approx
\exp( - t )$.  Effectively this approximation corresponds to the short
time expansion $\tilde G_{22} \sim 1/(z-\mel{\hat{Q}_{1}(0)}{{\cal
L}}{\hat{Q}_{1}(0)})$.  Even more remarkable is the clear emergence of
a plateau in the neighbor-pair autocorrelation function as $c$
decreases and the system becomes ``glassy'', yielding a two-step
relaxation time profile similar to that observed for the dynamic
structure factor at microscopic length scales in simple glass-forming
systems.  In such systems, the onset of the plateau, generally called
the $\beta$-regime, is relatively insensitive to temperature and is
often associated with the phenomenon of dynamic caging in dense fluid
systems.  In this regime, fluid particles typically oscillate in the
traps formed by their immediate neighbors and little relaxation of the
system occurs.  This behavior typically continues until a typical time
scale, known as the $\alpha$-regime, is reached in which particle
cages are temporarily broken.  This $\alpha$ time scale is strongly
temperature dependent and scales with the overall relaxation time of
the system.  Interestingly, similar behavior is observed in
$G^{(2)}_{22}(t)$ of the East model: There is an initially rapid decay
(with time scale $t \sim 1$) at which point a plateau appears.  The
plateau typically extends to times corresponding to the average
relaxation time $\tau$ of the spin autocorrelation function.  However,
unlike simple liquid systems, the \emph{height} of the plateau is
strongly temperature dependent, occurring roughly at value of $c$.  In
the East model, one can interpret the emergence of the plateau as
arising from a kind of effective dynamic caging of the pair spin
variable $n_i n_{i+1}$ that occurs when $n_{i+1}=1$.  When the right
neighbor of a given spin $i$ is up, the spin $n_i$ can oscillate
between values of $1$ and $0$ for extended periods of time,
corresponding to a kind of vibration in a cage.  This behavior will
persist until the spin $i+1$ flips, which typically will occur at
times $t \sim \tau$.  Furthermore, the probability of finding such a
caged system scales with the likelihood of finding an up-spin in
equilibrium, $c$.

The two-step relaxation of $G_{22}(t)$ was also found numerically by
Wu and Cao\cite{WuCao04} (who refer to this quantity $\mathsf C_2$).
Wu and Cao showed that the relaxation can be described with a
stretched exponential behavior at long times. From the numerical
analysis of our theoretical expressions, we find that the
parameter-free \emph{theoretical} relaxation profile is also
well-described by a stretched exponential with the same stretching
exponential $\beta \approx 0.6$ found in the analysis of the spin
autocorrelation function $C(t)$.

As can be seen from Fig.~\eqref{fourPoint}, the spin fluctuations
$\hat{n}_i$ do not behave as Gaussian random variables at all time
scales and for all values of $c$, unlike their counterpart, the
Fourier components of the mass density, in simple liquid systems.  It
can also be observed that the decay of the spin fluctuations is slower
than that predicted for a system exhibiting Gaussian statistics for
all times at high values of $c$.  As $c$ drops below $0.5$, the decay
becomes \emph{faster} than Gaussian at short times but slower than
Gaussian at long times.  The fact that $c=0.5$ is special can be seen
from a short time expansion of $\Delta (t)$:
\begin{eqnarray*}
 \Delta (t) &=&
 \mel{\hat{Q}_1(0)}{ e^{{\cal L} t} }{\hat{Q}_1(0)} - \mel{\hat{Q}_0}{
 e^{ {\cal L} t}}{ \hat{Q}_0}^2 \\
 &=&
 1 +  t \mel{\hat{Q}_1(0)}{ {\cal L}}{ \hat{Q}_1(0)} +  \frac{\,t^2}{2} \mel{\hat{Q}_1(0)}{{\cal L}^2}{ \hat{Q}_1(0)} \\
& &
 - \left[  1 + t \mel{\hat{Q}_0}{ {\cal L}^2}{ \hat{Q}_0} + \frac{\,t^2}{2}  \mel{\hat{Q}_0}{ {\cal L}^2 }{\hat{Q}_0} \right]^2
+O(t^3).
\end{eqnarray*}
Using the rules elaborated in Sec.~\ref{singlegap}, all quantities
appearing above are easily evaluated to reveal the exact result:
\begin{equation}
  \Delta (t) = (2c - 1) t  \left[1 - (2c+1) \frac{t}{2}\right] + O(t^3),
\end{equation}
from which the sign change for $c=0.5$ is explicitly evident at short times. 

One can also note in Fig.~\eqref{fourPoint} that the maximum positive
deviation from Gaussian behavior (i.e. slower than Gaussian) occurs at
a time which scales roughly with the average relaxation time $\tau$.

\section{\label{discussion}Discussion}

In this paper, the East model --- a linear kinetically constrained
spin model which is statically structureless --- was studied
theoretically taking the domains of down-spins as a starting
point. The constraints in the model lead to a very slow spin
relaxation for low up-spin density $c$ because of the existence of
these down-spin domains, of which each spin has to flip at least once
before a spin on the left of the domain can relax. Such highly
cooperative, hierarchical events driving the relaxation mimic
heterogeneous behavior in glasses.

The way the down-spin domains were taken into account was by using
them in the construction of a basis which is complete on the relevant
ergodic component.  In the complete domain basis, the theory is
formally exact, but the basis needs to be truncated to get explicit
results. In this truncation, one only limits how many simultaneous
domains are included without restricting the possible sizes of those
domains.  When we restricted ourselves to a single domain description,
an exact result for the single spin time correlation function $C(t)$
[$C^{(1)}$] was obtained which gives a good quantitative description
for $c$ larger than about $0.5$.  An extension including neighboring
domains led to an exact expression [$C^{(2)}$] which described the
slow, glassy behavior correctly down to $c\approx 0.3$. A general
procedure was outlined to obtain further approximations.

The main advantages of our approach over others are that a) it gives
explicit analytical results without fitting parameters, b) it requires
neither an arbitrary closure for the memory kernel nor the
construction of an irreducible memory kernel such as in mode coupling
theories and c) nonetheless, it described low $c$ behavior equally
well as these mode coupling theories. The explanation for this power
is that domains of all sizes are included.

At a given level of truncation, the matrix approach outlined here
allows analytical results for the spin autocorrelation function to be
obtained.  Armed with these results, it is possible to assess the
effect of truncation the multi-domain basis by evaluating approximate
expressions for the ``self-energy'' terms, as was done in
Sec.~\ref{gapbasis}.  One can then examine the time scale at which the
higher domain corrections become important and their magnitude for a
given value of $c$.  Such information is useful in examining dynamical
scaling relations\cite{RitortSollich03,Ludo04}.

The matrix approach is also well-suited for examining higher-order
correlation functions, such as the neighbor-pair auto-correlation
function, that probe detailed aspects of the dynamics, as was shown in
section~\ref{higherorder}.

Our theory does not require an ansatz for a closure relation between
the memory kernel and the correlation function, yet it does have the
\emph{structure} of a mode coupling theory. First of all, the theory,
derived using a projection operator formalism, yields a basis set very
similar to the multi-linear set in the theory of Oppenheim {\em et
al.}  Secondly, successive truncations of the set are like including
only linear modes, or also bilinear modes, or also tri-linear ones,
etc., again very similar to mode coupling theories for fluids. Finally
and perhaps most strikingly, without assuming a closure relation, a
self-consistent equation emerges for part of the result,
i.e., for $\tilde\gamma^{(1)}$ in Eq.~\eqref{gammaeq} of
Sec.~\ref{singlegap} and for $\tilde\gamma^{(2)}$ in
Eq.~\eqref{tildegammaeq} of Sec.~\ref{extendedgap}. Thus, in a sense,
the correct closure relation follows unambiguously from the theory
rather than being assumed. Perhaps this is an indication why mode
coupling theories can work, at least in some range of $c$, if the
closure relation is well chosen. However, as the difference between
the closure for $\tilde\gamma^{(1)}$ and $\tilde\gamma^{(2)}$ shows,
the required closure depends on how low $c$ is. The closure can also
become ``hierarchical'', in the sense that $\tilde\gamma^{(2)}$
depends on~$\tilde\eta_2$, which itself satisfies a self-consistent
equation.

A natural question is how adaptable is the matrix approach outlined
here for other conditions of spin facilitation, such as the
Frederickson-Andersen\cite{FredericksonAndersen84,FredericksonAndersen85}
(FA) model, higher-dimensions and other types of lattices.  The
extension to the FA model involves extending the one-domain basis set
to include domains on \emph{both} sides of the targeted spin and
involves slightly more complicated matrix algebra than that presented
here\cite{vanZonFA04}.  For the FA model for which self-consistent
closure schemes in the context of mode-coupling theory appear to work
quite well\cite{Szamel04}, quite good quantitative agreement can be
obtained with the simple single-domain basis set.  Extensions to
include multiple domains can be carried out numerically for the FA
model as well as other generic models.  In addition,
higher-dimensional models can also be tackled in a numerical fashion
using finite basis set representations, provided the basis sets
include domains that are sufficiently large.  Although finite matrix
representations are always bound to give the incorrect long time
asymptotic behavior for systems exhibiting stretched-exponential
profiles, the short and intermediate time behavior can be reproduced
with great accuracy.

It is conceivable that the complete basis set presented in
Sec.~\ref{extendedgap} has a deeper structure that could be exploited
for the description for $c\to0$.  Also, the domain basis might be used
to describe the response of the East model to a sudden ``quench'' to
low $c$ values.  Work on these issues is in progress.

Finally, our approach shows how important it is to \emph{first}
identify the ``slow modes'' of a system, in this case the down-spin
domains, \emph{before} embarking on a mode coupling-like description
of the long time behavior of correlation functions.

\begin{acknowledgments}

This work was supported by a grant from the Natural Science and
Engineering Research Council of Canada.  For the final part of this
project, R.V.Z. acknowledges the support of the Office of Basic Energy
Sciences of the US Department of Energy, under grant No.\
DE-FG-02-88-ER13847. J.S. would also like to thank Walter Kob and the
Laboratoire des Verres in Montpellier, France for their hospitality
and the CNRS for additional funding during the final stages of this
work.

\end{acknowledgments}

\appendix*

\section{\label{appA}Self-energy matrix in the two-domain basis}

From the expression for the self-energy matrix in the two-domain
approximation,
\[
\tilde{\Sigma}_{11}(z) \approx \cfrac{\mathsf M_{12}\,\mathsf
        M_{12}^\dagger}{z\identity-\mathsf M_{22}},
\]
it is clear that we must evaluate matrices such as $\mathsf M_{12} =
\mel{\hat Q_1(l)}{\mathcal L}{\hat Q_2(l_1, l_2)}$.  The double
indices on $\hat Q_2(l_1,l_2)$ tend to make the algebra somewhat less
transparent than in Sec.~\ref{singlegap}, and it turns out that the
self-energy matrix can be evaluated more easily by splitting up the
set $\hat Q_2$ into two-domain variables for which $l_2=0$ and those
for which $l_2>0$, by defining
\begin{eqnarray}
   \hat R(k) &=& \hat Q_2(k,0)\\
   \hat S(l_1,l_2) &=& \hat Q_2(l_1,l_2+1)
\end{eqnarray}
and likewise for unnormalized versions, as indicated in
Table~\ref{Bdiagrams}.  The matrix $\mathsf M_{12}$ then takes on the
form $\mathsf M_{12} = \left[ \mathsf M_{QR} , \mathsf M_{QS} \right]$,
where
\begin{eqnarray}
  \mathsf M_{QR} &=&
    \mel{\hat Q_1}{\mathcal L}{\hat R}
  \label{MEBA}
\\
  \mathsf M_{QS} &=&
    \mel{\hat Q_1}{\mathcal L}{\hat S}
  \label{MECA}
\end{eqnarray}
and $\mathsf M_{22}$ is written in the block form
\begin{equation}
  \mathsf M_{22} = \left[ \begin{matrix}
    \mathsf M_{RR} & \mathsf M_{RS}\\&&\\
    \mathsf M_{RS}^\dagger & \mathsf M_{SS}
    \end{matrix}
    \right]
\label{splitME}
\end{equation}
where $\mathsf M_{RR}$,  $\mathsf M_{RS}$ and $\mathsf M_{SS}$ are
\begin{eqnarray}
  \mathsf M_{RR} &=& \mel{\hat R}{\mathcal L}{\hat R}
\label{MEBB}
\\
  \mathsf M_{RS} &=& \mel{\hat R}{\mathcal L}{\hat S}
\label{MEBC}
\\
  \mathsf M_{SS} &=& \mel{\hat S}{\mathcal L}{\hat S},
\label{MECC}
\end{eqnarray}
and where the notation that $\hat R$ or $\hat S$ without any
argument denotes the column or row vector composed of all $\hat
R(k)$ or $\hat S(k_1,k_2)$ respectively.

The matrix $\mathsf M_{QS}$ is in fact zero, so the matrix
self-energy $\tilde{\mathsf\Sigma}_{11}$ is
\begin{equation}
   \tilde\Sigma_{11} =
\mathsf M_{QR} \Big[z\identity - \mathsf M_{22} \big]^{-1}_{RR} \mathsf M_{QR}^{\dagger}.
\end{equation}
Using the matrix equality ~\eqref{splitinverse}, this expression can
be re-written as
\begin{eqnarray}
   \tilde\Sigma_{11} &=& 
\mathsf M_{QR} \Bigg[z\identity - \mathsf M_{RR}
-\cfrac{\mathsf M_{RS} \, \mathsf M_{RS}^\dagger}{z\identity-\mathsf M_{SS}}\Bigg]^{-1}\mathsf M_{QR}^\dagger
\nonumber\\
&=&
\cfrac{\mathsf M_{QR}\,\mathsf M_{QR}^\dagger}{
z\identity - \mathsf M_{RR}
-\cfrac{\mathsf M_{RS}\, \mathsf M_{RS}^\dagger}{z\identity-\mathsf
  M_{SS}}}
.
\label{SigmaAA}
\\&&\nonumber
\end{eqnarray}

The explicit calculation of all the matrix elements appearing in
Eq.~\eqref{SigmaAA} proceeds as follows: We start with $\mathsf M_{QR}$
defined in Eq.~\eqref{MEBA}.  Combining the diagrams of $\mathcal
LQ_1(k)$ in Eqs.~\eqref{LQ10} and \eqref{LQ1k} with the diagrams of
$R(k')=Q_2(k',0)$ in Table~\ref{Bdiagrams} yields
\begin{eqnarray*}
  \mel{Q_1(k)}{\mathcal L}{R(k-1)} &=&
(1-c)\diagram{9}{
        \setO\txt{k-1}\cbar\cbar\cbar\cbar\setO\setX \nxt
        \setO\cbar\txt{}\cbar\cbar\cbar\setX\setO
        }
=
c^3 (1-c)^{k+3}
\\
  \mel{Q_1(k)}{\mathcal L}{R(k)} &=&
        -\diagram{9}{
        \setO\cbar\cbar\txt{k}\cbar\cbar\setO\setX \nxt
        \setO\cbar\cbar\txt{}\cbar\cbar\setX\setO
        }
=- c^3 (1-c)^{k+3}
\end{eqnarray*}
while all other $\mel{Q_1(k)}{\mathcal L}{R(k')}$ are zero. By similar
diagrammatic means, one finds $\mel{S(l_1,l_2)}{\mathcal L}{Q_1(k)}=0$
--- so that $\mathsf M_{QS}$ in Eq.~\eqref{MECA} is indeed zero as we
anticipated above. Also $\mel{R(k)}{\mathcal
L}{Q_0}=\mel{S(l_1,l_2)}{\mathcal L}{Q_0}=0$, confirming that $\mathsf
M_{02}=0$.  Using Eqs.~\eqref{MEBA}, as well as \eqref{z1def} and
\eqref{theno} gives
\begin{eqnarray}
  \mathsf M_{QR} =
  \left[\begin{matrix}
  -(1-c)c^{1/2}&0\\
  (1-c)^{3/2}c^{1/2} &\ddots&\ddots\\
  &\ddots
  \end{matrix}\right].
\label{MEBAresult}
\end{eqnarray}

Next we will determine $\mathsf M_{RR}$ defined in \eqref{MEBB}. For
this we need the diagrams of $\mathcal L R(k)$:
\begin{eqnarray}
\mathcal L  R(0) &=& - \dline{5}{\setO\setX\setO\setX} - (1-c) \dline{4}{\setO\setO\setX}
 - \dline{4}{\setO\setX\setO}
\label{Rnulresult}
\\
\mathcal L  R(k\geq1)
&=&   -   \dline{10}{\setO\cbar\cbar\txt{k}\cbar\cbar\setX\setO\setX}
             - (1-c)
                 \dline{9}{\setO\cbar\cbar\txt{k}\cbar\cbar\setO\setX}
                 +\dline{8}{\setO\txt{k-1}\cbar\cbar\cbar\cbar\setO\setX\setO}
\nonumber\\&&
\label{Rkresult}
\end{eqnarray}
Combining these diagrams with those of $R(k')$ in Table~\ref{Bdiagrams}
yields
\begin{eqnarray}
  \mel{R(0)}{\mathcal L}{R(0)} &=&
     -\diagram{5}{
          \setO\setX\setO\setX\nxt
          \setO\setX\setO
          }
        -(1-c)\diagram{4}{
          \setO\setO\setX\nxt
          \setO\setX\setO
        }
        -\diagram{4}{
          \setO\setX\setO\nxt
          \setO\setX\setO
          }
\nonumber\\
&=&
- c^3(1-c)^2(2-c+c^2) 
\\
  \mel{R(k)}{\mathcal L}{R(k)} &=&
-     \diagram{10}{
          \setO\cbar\cbar\txt{k}\cbar\cbar\setX\setO\setX\nxt
          \setO\cbar\cbar\cbar\cbar\cbar\setX\setO
      }
 -(1-c)\diagram{9}{
          \setO\cbar\cbar\txt{k}\cbar\cbar\setO\setX\nxt
          \setO\cbar\cbar\cbar\cbar\cbar\setX\setO
       }
     +\diagram{10}{
          \setO\txt{k-1}\cbar\cbar\cbar\cbar\setO\setX\setO\nxt
          \setO\cbar\cbar\cbar\cbar\cbar\cbar\setX\setO
       }
\nonumber
\\&=&
- c^3 (1-c)^{k+2}(1+c^2)  
\hfill\text{ if } k\geq1
\end{eqnarray}
while $  \mel{R(k)}{\mathcal L}{R(k')}=0$ if $k\neq k'$.
Using also Eqs.~\eqref{theno} and \eqref{unno}, one finds
\begin{equation}
\left[\mathsf  M_{RR}\right]_{kk'} = -(2-c+c^2)\delta_{k0}\delta_{kk'} -
(1+c^2)(1-\delta_{k0})\delta_{kk'},
\label{MRR}
\end{equation}

Next to determine is $\mathsf M_{RS}$. Combining the diagrams of $S$ in
Table \ref{Bdiagrams} with those of $\mathcal LR$ in
Eqs.~\eqref{Rnulresult} and \eqref{Rkresult}, we see that many
elements of $\mathsf M_{RS}$ are zero, while the nonzero ones are
restricted to
\begin{equation}
  \mel{S(k,0)}{\mathcal L}{R(k)} = - 
        \diagram{10}{
          \setO\cbar\cbar\txt{k}\cbar\cbar\setX\setO\setX\nxt
          \setO\cbar\cbar\cbar\cbar\cbar\setX\cbar\setO
        }
= c^4(1-c)^{k+3}
\end{equation}
for all $k\geq0$. Using Eqs.~\eqref{theno} and \eqref{MEBC}, we find 
\begin{equation}
  \left[\mathsf M_{RS}\right]_{kl_1l_2}
 = c(1-c)^{1/2}\delta_{l_1k}\delta_{l_20}
\label{60}
\end{equation}

The final matrix to determine is $\mathsf M_{SS}$, for which we require
$\mathcal LS(l_1,l_2)$:
\begin{eqnarray*}
  \mathcal L S(0,l_2) &=&
-
\dline{10}{\setO\setX\txt{l_2+1}\cbar\cbar\cbar\cbar\cbar\setO}
+(1-c)
\dline{8}{\setO\setX\cbar\txt{l_2}\cbar\setO\setX}
-
\dline{10}{\setO\setX\txt{l_2+1}\cbar\cbar\cbar\cbar\cbar\setO\setX}
\\
  \mathcal L S(l_1\geq 1,l_2) &=&
\dline{18}{\setO\txt{l_1-1}\cbar\cbar\cbar\cbar\cbar\setO\setX\cbar\txt{l_2+1}\cbar\cbar\cbar\cbar\cbar\setO}
+(1-c)
\dline{11}{\setO\cbar\txt{l_1}\cbar\setX\cbar\txt{l_2}\cbar\setO\setX}
\\&&
-
\dline{15}{\setO\cbar\txt{l_1}\cbar\setX\cbar\txt{l_2+1}\cbar\cbar\cbar\cbar\cbar\setO\setX}
.
\end{eqnarray*}
Combining with the diagrams of $S$ in Table~\ref{Bdiagrams}, one finds
\begin{eqnarray*}
  \mel{S(0,l_2)}{\mathcal L}{S(0,l_2)} &=& -(1+2c-c^2)c^3(1-c)^{3+l_2}\\
  \mel{S(0,l_2+1)}{\mathcal L}{S(0,l_2)} &=& c^4(1-c)^{l_2+4}\\
  \mel{S(l_1\geq1,l_2)}{\mathcal L}{S(l_1,l_2)} &=&
  -(3-c)c^4(1-c)^{l_1+l_2+3}\\
  \mel{S(l_1\geq1,l_2+1)}{\mathcal L}{S(l_1,l_2)} &=&
  c^4(1-c)^{l_1+l_2+4},
\end{eqnarray*}
while all other $\mel{S}{\mathcal L}{S}$ are zero, Combining with
Eqs.~\eqref{theno} and \eqref{MECC}, one gets
\begin{eqnarray}
\left[\mathsf  M_{SS}\right]_{l_1l_2l_1'l_2'} 
&=& \delta_{l_1l_1'}\big[
  (\delta_{l_2'l_2+1}+\delta_{l_2'l_2-1})c(1-c)^{1/2}
\nonumber\\
&&\qquad  -\delta_{l_2'l_2}\delta_{l_10}(1+2c-c^2)
\nonumber\\
&&\qquad
  -\delta_{l_2'l_2}(1-\delta_{l_10})c(3-c)
\big].
\end{eqnarray}

In view of Eqs.~\eqref{SigmaAA} and \eqref{60}, we need the
$l_2=0,l_2'=0$ component of the inverse of $z\identity-\mathsf
M_{SS}$. This matrix is diagonal in $l_1$ and $l_1'$ and
tri-diagonal in $l_2$ and $l_2'$ for fixed $l_1$ and
$l_1'$. Thus, we can use Eq.~\eqref{contfrac} to write
\begin{eqnarray}
  [z\identity-\mathsf M_{SS}]^{-1}_{l_1,0;l_1;0} &=& 
  \frac{\delta_{l_10}}{\tilde a_1-\tilde\varepsilon_{1}}
+  \frac{1-\delta_{l_10}}{\tilde a_2-\tilde\varepsilon_{2}},
\label{MSS}
\end{eqnarray}
where 
\begin{subequations}
\label{aj}
\begin{eqnarray}
  \tilde a_1 &=& z+1+2c-c^2
\\
  \tilde a_2 &=& z+c(3-c).
\end{eqnarray}
\end{subequations}
and $\tilde\varepsilon_{1}$ and $\tilde \varepsilon_{2}$ result from
the repeating part of the continued fraction that results from
applying Eq.~\eqref{contfrac}. Similar to $\tilde\gamma^{(1)}$ in
Eq.~\eqref{gammaeq} in section \ref{singlegap}, they satisfy
\begin{equation}
\tilde\varepsilon_{j} = {c^2(1-c)}/({\tilde a_j-\tilde\varepsilon_{j}}).
\label{epsself}
\end{equation}
With the requirement that they go as $1/z$ for large $z$, the
solutions are
\begin{eqnarray}
  \tilde\varepsilon_{j} &=& 
\frac12\left[\tilde a_j-\sqrt{\tilde a_j^2-4c^2(1-c)}\right].
\label{oldej}
\end{eqnarray}
The subexpression $\mathsf M_{RS}[z\identity-\mathsf M_{SS}]^{-1}\mathsf
M_{RS}^\dagger$ in Eq.~\eqref{SigmaAA} now becomes, using
Eq.~\eqref{60}, \eqref{MSS}, \eqref{aj} and \eqref{epsself}
\begin{equation}
  \left[\frac{\mathsf M_{RS}\,\mathsf M_{RS}^\dagger}{z\identity-\mathsf M_{SS}}\right]_{kk'}
= \delta_{kk'}\left[ \tilde\varepsilon_1\delta_{k0}+
 \tilde\varepsilon_2(1-\delta_{k0})
\right]
\end{equation}
Since this matrix and $\mathsf M_{SS}$ in Eq.~\eqref{MRR} are diagonal,
the inverse of $(z\identity-M_{RR}-\mathsf M_{RS}[z\identity-\mathsf
M_{SS}]^{-1}\mathsf M_{RS}^\dagger)$ is simply
\begin{eqnarray*}
&&
\left[z\identity-\mathsf  M_{RR}-\cfrac{\mathsf M_{RS}\,\mathsf
    M_{RS}^\dagger}{z\identity-\mathsf M_{SS}}\right]^{-1}_{kk'} 
\\&&\quad
=
\frac{\delta_{k0}\delta_{kk'}}{z+2-c+c^2-\tilde\varepsilon_{1}} +
\frac{(1-\delta_{k0})\delta_{kk'}}{z+1+c^2-\tilde\varepsilon_{2}}
\\&&\quad
=
\frac{\delta_{kk'}}{1-c}
\left[
\frac{\delta_{k0}}{1-2c+c^2/\tilde\varepsilon_1} +
\frac{1-\delta_{k0}}{1-2c+c^2/\tilde\varepsilon_2}
\right]
,
\end{eqnarray*}
where in the last equality we used again Eq.~\eqref{epsself}. Given
this last form, we can shorten many equations by using the expressions
$\tilde\eta_j=c(1-c)/(1-2c+c^2/\tilde\varepsilon_j)$, which are
explicitly given by
\begin{equation}
\tilde\eta_j
=
\cfrac{c(1-c)}{1-2c+\cfrac{2c^2}{\tilde a_j-\sqrt{\tilde a_j^2-4c^2(1-c)}}}.
\label{ej}
\end{equation}
and in terms of which we have
\[
\left[z\identity-\mathsf  M_{RR}-\cfrac{\mathsf M_{RS}\,\mathsf
    M_{RS}^\dagger}{z\identity-\mathsf M_{SS}}\right]^{-1}_{kk'} 
=
\frac{\delta_{kk'}[
\tilde\eta_1\delta_{k0}+\tilde\eta_2(1-\delta_{k0})]
}{c(1-c)^2}
.
\]
Inserting this result in the expression for the self-energy matrix in
Eq.~\eqref{SigmaAA}, one obtains the result presented in
Eq.~\eqref{Sigma}.

\end{document}